\documentclass[english,amssymb,amsmath,showpacs,twocolumn,noshowpacs,prl]{revtex4}
\usepackage[english]{babel}
\usepackage[utf8]{inputenc}
\usepackage[colorlinks,pdfpagelabels,pdfstartview = FitH,bookmarksopen = true,bookmarksnumbered = true,linkcolor = black,plainpages = false,hypertexnames = false,citecolor = black] {hyperref}
\usepackage{mathtools}
\usepackage{textcomp}
\usepackage{overpic}

\begin{document}
	
	\title{Exciton-Scattering-Induced Dephasing in Two-Dimensional Semiconductors}
	
	\author{Florian Katsch}
	\author{Malte Selig}
	\author{Andreas Knorr}	
	\affiliation{Institut f\"ur Theoretische Physik, Nichtlineare Optik und Quantenelektronik, Technische Universit\"at Berlin, 10623 Berlin, Germany}
	
	\begin{abstract}
		Enhanced Coulomb interactions in monolayer transition metal dichalcogenides cause tightly bound electron-hole pairs (excitons) which dominate their linear and nonlinear optical response.
		The latter includes bleaching, energy renormalizations, and higher-order Coulomb correlation effects like biexcitons and excitation-induced dephasing (EID).
		While the first three are extensively studied, no theoretical footing for EID in exciton dominated semiconductors is available so far.
		In this study, we present microscopic calculations based on excitonic Heisenberg equations of motion and identify the coupling of optically pumped excitons to exciton-exciton scattering continua as the leading mechanism responsible for an optical power dependent linewidth broadening (EID) and sideband formation.
		Performing time-, \mbox{momentum-,} and energy-resolved simulations, we quantitatively evaluate the EID for the most common monolayer  transition metal dichalcogenides and find an excellent agreement with recent experiments.
	\end{abstract}
	
	\maketitle

	The widely investigated nonlinear optical response performed on III-V quantum wells involves bleaching and energy renormalizations of the optical transitions, as well as non-perturbative electronic correlation effects like biexciton formation and a spectral linewidth increase with rising excitation density.
	The latter, often referred to as excitation-induced dephasing (EID) \cite{schultheis1986ultrafast,honold1989collision,wang1993transient,wang1994transient,hu1994excitation,rappen1994polarization,wagner1997coherent,wagner1999interaction,li2006many,moody2011exciton,nardin2014coherent}, originates from the simultaneous optical excitation of excitons and incoherent electron-hole plasma interacting via frequency-dependent screened Coulomb interactions with nonzero imaginary part as demonstrated within a kinetic Markovian Boltzmann-like scattering theory \cite{wang1994transient}.

	\begin{figure}
		\centering
		\begin{overpic}[width=0.95\columnwidth]{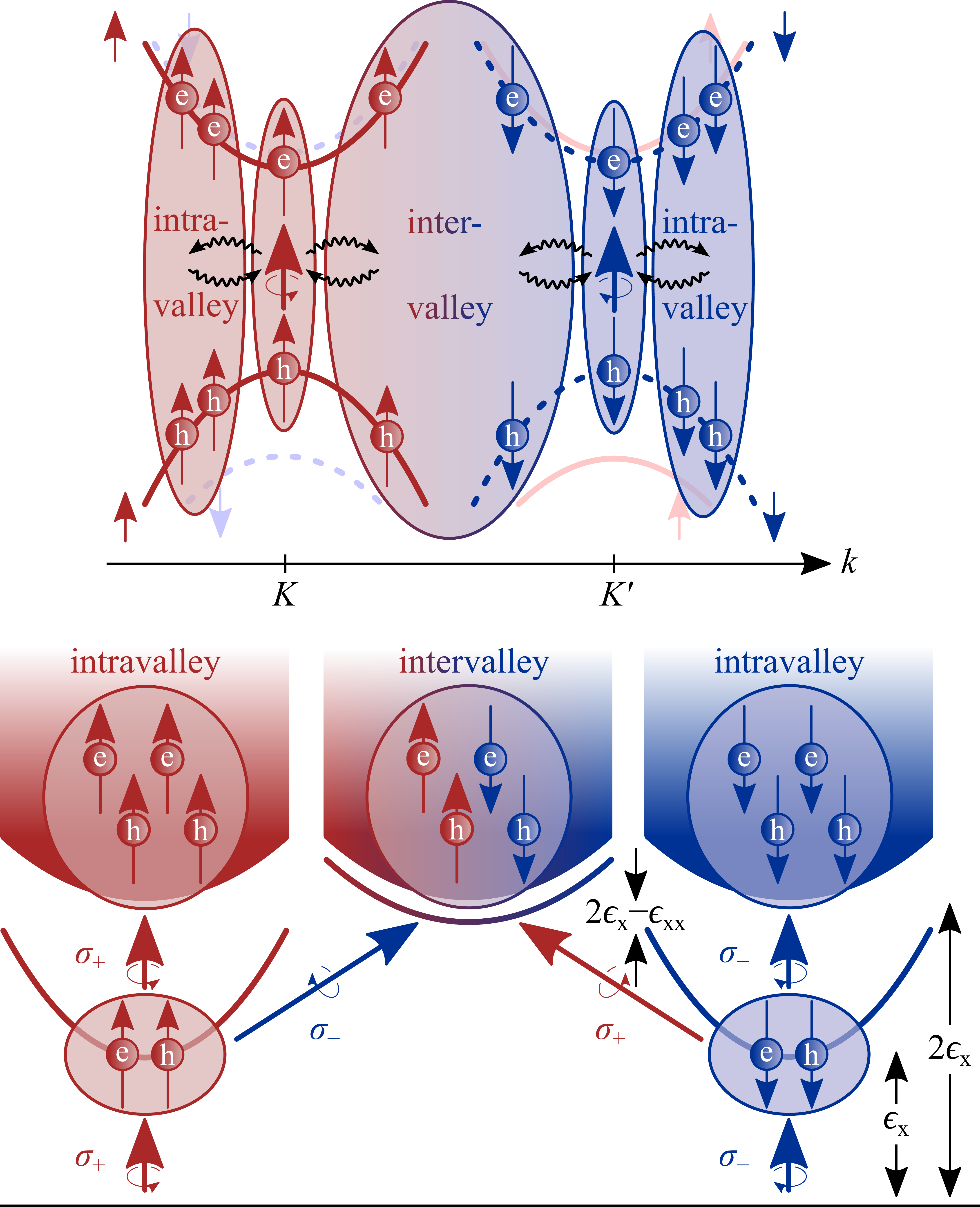}
			\put(0,97.5){(a)}
			\put(0,47){(b)}
		\end{overpic}
		\caption{\textbf{Excitonic correlations in monolayer TMDCs.}
			(a)~$\sigma_{+(-)}$~polarized light introduces bound electron-hole pairs (excitons) near the corner $K^{(\prime)}$ of the hexagonal Brillouin zone.
			In the nonlinear regime, optically addressable excitons interact via Coulomb interactions with two-electron and two-hole Coulomb intravalley and intervalley correlations.
			(b)~Optically addressable excitons with energy~$\epsilon_\text{x}$ couple to intravalley exciton-exciton scattering continua setting in at $2\epsilon_\text{x}$ excited by $\sigma_{+}$ or $\sigma_{-}$~polarized light as well as intervalley biexcitons separated by $2\epsilon_\text{x}$$-$$\epsilon_{\text{xx}}$ from the associated intervalley exciton-exciton scattering states addressed by the simultaneous excitation with $\sigma_{+}$ and $\sigma_{-}$~polarized light.
		}
		\label{Bild-Schema}
	\end{figure}
	However, recent advance in growth technology enabled the fabrication of purely two-dimensional semiconductors \cite{novoselov2005two} including monolayers of transition metal dichalcogenides (TMDCs) \cite{mak2010atomically,splendiani2010emerging}.
	The reduced dimensionality and strong Coulomb interactions enabled the investigation of nearly pure excitonic systems with tightly bound electron-hole pairs (excitons) separated by hundreds of meV from the free particle band gap \cite{berkelbach2013theory,chernikov2014exciton}.
	Hence, intense many-body effects make atomically thin TMDCs to ideal candidates for studying a completely new type of EID originating from exciton-exciton scattering in the absence of free electron-hole plasma \cite{moody2015intrinsic,dey2016optical,mahmood2017observation,martin2018encapsulation,jakubczyk2019coherence,boule2020coherent}.
	Whereas	an enhancement of the EID in monolayer TMDCs by one order of magnitude compared to conventional semiconductor quantum wells was experimentally demonstrated \cite{moody2015intrinsic}, a profound theoretical understanding of underlying mechanisms is still missing.
	In this work, we provide a microscopic description of EID of excitons in two-dimensional semiconductors.
	Our approach is based on the excitonic Heisenberg equation of motion formalism \cite{ivanov1993self,katsch2018theory} including the Coulomb coupling of optically pumped excitons to (i)~other excitons, (ii)~biexcitons \cite{kylanpaa2015binding,mayers2015binding,zhang2015excited,kidd2016binding,mostaani2017diffusion,kezerashvili2017trion,szyniszewski2017binding,van2018excitons,steinhoff2018biexciton,kuhn2019combined} detectable below the energetically lowest excitons (binding energy $2\epsilon_\text{x}$$-$$\epsilon_\text{xx}$) \cite{sie2015intervalley,aleithan2016broadband,sie2016observation,hao2017neutral}, and (iii)~unbound solutions of the two-electron and two-hole Schr\"odinger equation, cf. Fig.~\ref{Bild-Schema}.
	The latter describe continuous Coulomb correlated exciton-exciton scattering states setting in at twice the exciton energy $2\epsilon_\text{x}$ \cite{axt1998exciton,bartels1998identification,kwong2001third,kwong2001evidence,takayama2002t,voss2002biexcitonic,schumacher2005coherent,schumacher2006coherent,voss2006coherent,schumacher2007influence,schafer2013semiconductor,katsch2020theory} and provide new scattering channels for optically addressable excitons.
	Based on time-, momentum-, and energy-resolved simulations, we quantitatively identify the coupling of excitons to exciton-exciton scattering continua as the leading mechanism for the excitation power dependent linewidth broadening.
	We calculate a linear increase of the exciton linewidth with rising excitation power for the most common TMDCs and find an excellent agreement with available experiments \cite{moody2015intrinsic,martin2018encapsulation,boule2020coherent}.
	Besides, we report excitation power dependent asymmetric lineshapes with sidebands on the high energy sides of exciton resonances, as expected from the fluctuation-dissipation theorem, originating from the non-Markovian interaction of excitons with exciton-exciton scattering continua.
\textit{Excitonic description}:
	Bound electron-hole pairs are described by first solving the Wannier equation, which provides the exciton energies~$\epsilon_{\text{x},\lambda}$ as well as wave functions~$\varphi_{\lambda,\textbf{\textit{q}}}$ \cite{kira2006many}, and subsequently calculating the dynamics of the excitonic interband transitions $P_{\lambda}(t)$ \cite{katsch2018theory}.
	$\lambda$ as a compound index incorporates the exciton state, the high-symmetry point, and the spins of involved electrons and holes.
	The Heisenberg equation of motion for optically addressable excitonic transitions~$P_{\lambda}$ reads, cf. Supplementary Section~1:
	\begin{align}
	& \left(\hbar\partial_t + \gamma_0 - \text{i} \epsilon_{\text{x},\lambda_1} \right) P^{\phantom{*}}_{\lambda_1} \notag\\
	& = -\text{i} \, {d}_{\lambda_1} {E}_{\sigma_\pm}(t) +\text{i} \, \sum_{\lambda_2,\lambda_3} \hat{{d}}_{\lambda_1,\lambda_2,\lambda_3} {E}_{\sigma_\pm}(t) \ P^{\phantom{*}}_{\lambda_2} P^*_{\lambda_3} \notag \\
	& \hspace{4mm} + \text{i} \sum_{\lambda_2} V_{\lambda_1,\lambda_2} \ P^{\phantom{*}}_{\lambda_2}
	+ \text{i} \sum_{\lambda_2,\lambda_3,\lambda_4} \hat{V}_{\lambda_1,\lambda_2,\lambda_3,\lambda_4} \ P^{\phantom{*}}_{\lambda_2} P^{\phantom{*}}_{\lambda_3} P^*_{\lambda_4} \notag \\
	& \hspace{4mm} + \text{i}\sum_{\lambda_2,\eta} {W}_{\lambda_1,\lambda_2,\eta} \ B^{\phantom{*}}_{\eta} P^*_{\lambda_2} .
	\label{eq:x-schematisch}
	\end{align}
	Eq.~\eqref{eq:x-schematisch} describes optical oscillations (energy $\epsilon_{\text{x},\lambda_1}$) damped by the microscopically calculated phonon-mediated dephasing rate $\gamma_0$ \cite{selig2016excitonic}, cf. Supplementary Section~2.
	$\sigma_{+(-)}$~circularly polarized light ${E}_{\sigma_{+(-)}}(t)$ drives $P_{\lambda_1}$ according to the optical selection rules \cite{xiao2012coupled} incorporated in the excitonic dipole matrix element~${d}_{\lambda_1}$ (first term in the second line).
	Self-consistently solving the TMDC Bloch equations and Maxwell's equations determines the radiative dephasing \cite{knorr1996theory}.
	Additionally, the exciton-light interaction involves Pauli blocking (second term in the second line).
	Coulomb scattering comprises linear and nonlinear Hartree--Fock interactions (third line) and the coupling of excitons to two-electron and two-hole Coulomb correlations~$B_{\eta}$ (fourth line) \cite{axt1994dynamics}.
	$\eta$ comprises the high-symmetry points and spins of the two electrons and two holes.
	Solving the \textit{intravalley} two-electron and two-hole Schrödinger equation for two electrons and two holes at the $K$~point accesses {intravalley} exciton-exciton scattering continua energetically above the exciton energy, cf. Fig.~\ref{Bild-Schema}~(b).
	Besides \textit{intervalley} exciton-exciton scattering continua, the {intervalley} four-particle eigenvalue problem with one electron and hole near the $K$~point and one electron and hole near the $K'$~point results in bound {intervalley} biexcitons.
	Biexcitons are observed energetically below the exciton resonance due to the attractive interaction of two virtual excitons with opposite spins near distinguished high-symmetry points.
	The Heisenberg equations of motion for biexcitons and exciton-exciton scattering states~$B_{\eta}$ describe damped ($\gamma_\text{xx}$) oscillations (energy ${\epsilon}_{\text{xx},\eta}$) self-consistently coupling to two excitons~$P_{\lambda_1} P_{\lambda_2}$ mediated by Coulomb interactions ($\hat{W}_{\eta,\lambda_1,\lambda_2}$) \cite{gamma}:
	\begin{align}
	& \left( \hbar \partial_t +\gamma_\text{xx} +\text{i} {\epsilon}_{\text{xx},\eta} \right) B^{\phantom{*}}_{\eta} = \text{i} \sum_{\lambda_1,\lambda_2} \hat{W}_{\eta,\lambda_1,\lambda_2} \ P^{\phantom{*}}_{\lambda_1} P^{\phantom{*}}_{\lambda_2} . \label{eq:BiX}
	\end{align}
	\begin{figure*}
		\centering
			\begin{overpic}[width=0.42\textwidth]{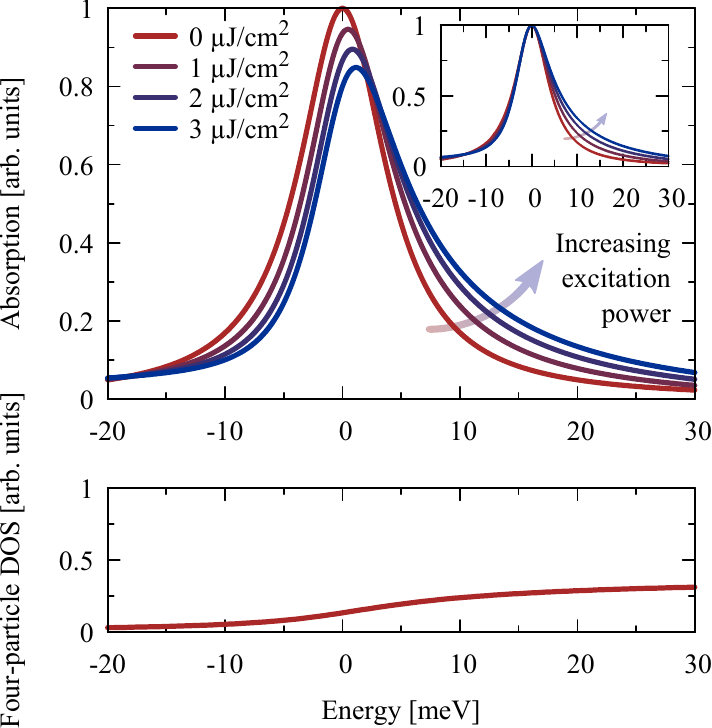}
				\put(-0.5,96.9){(a)}
				\put(23,60){\large$\circlearrowright$}
			\end{overpic}
			\hspace{10mm}
			\begin{overpic}[width=0.42\textwidth]{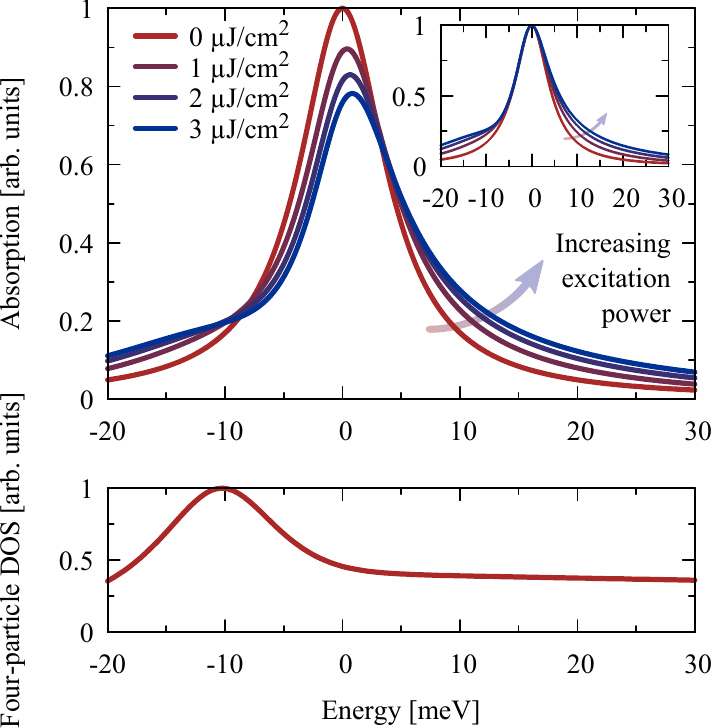}
				\put(-0.5,96.9){(b)}
				\put(23,60){\large$\uparrow$}
			\end{overpic}\vspace{8mm}
		
			\begin{overpic}[width=0.66\textwidth]{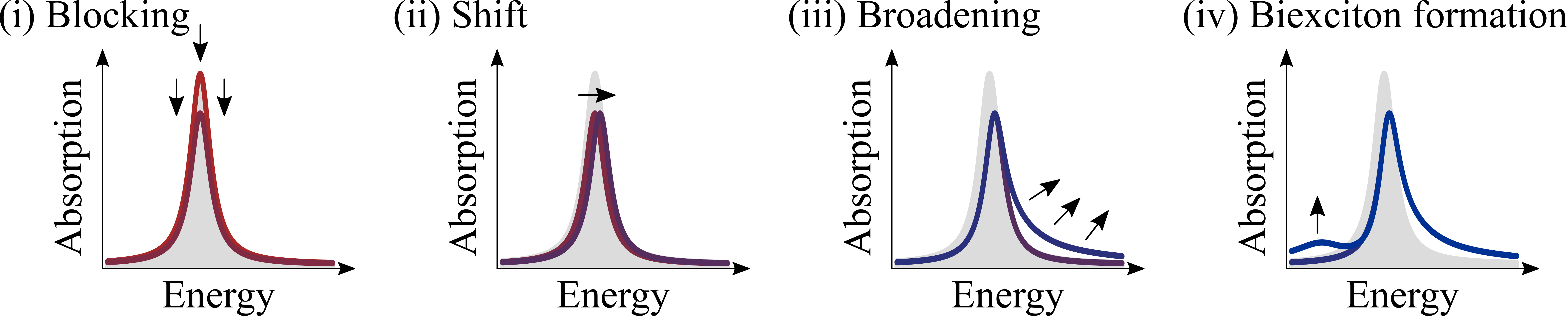}
				\put(-5,21){(c)}
			\end{overpic}
		\caption{\textbf{Nonlinear absorption of WSe$_2$.}
			(a,b)~Depicted is the nonlinear absorption spectrum of monolayer WSe$_2$ on a SiO$_2$ substrate at 10~K for different pump fluences.
			The energetically lowest optically addressable exciton is resonantly excited by (a) circularly ($\circlearrowright$) or (b) linearly ($\uparrow$) polarized, 40~fs pulses.
			The insets show the normalized and shifted absorption revealing non-Markovian sidebands on the high energy sides of the exciton resonances which become more pronounced for higher excitation densities.
			The four-particle (two-electron and two-hole) densities of states (DOS) for circularly and linearly polarized excitation are plotted below.
			(c)~Illustrated are the four effects observed in the microscopically calculated nonlinear absorption, namely (i)~a blocking of the absorption, (ii)~a shift of the resonance energy, (iii)~a line broadening originating from a sideband formation on the high energy side of the exciton resonance, and (iv) the formation of biexcitons for linearly polarized excitation.
			The shaded area represents the absorption for negligibly small excitation strength.
		}
		\label{Bild-Exzitonen}
	\end{figure*}
\textit{Results}:
	Self-consistently evaluating the coupled exciton, biexciton, and exciton-exciton scattering continua dynamics, Eqs.~\eqref{eq:x-schematisch} and \eqref{eq:BiX}, together with Maxwell's equations \cite{knorr1996theory} in an non-perturbative way accesses \cite{haase2000intensity} the nonlinear frequency-dependent absorption $\alpha \left(\omega\right)= 1 - T \left(\omega\right) - R \left(\omega\right)$ defined with respect to the transmission $T \left(\omega\right)$ and reflection $R \left(\omega\right)$ \cite{kira2006many}.
	Fig.~\ref{Bild-Exzitonen}~(a,b) shows the absorption of the energetically lowest optically addressable exciton for resonant excitation of WSe$_2$ on a SiO$_2$ substrate at 10~K with (a) circularly or (b) linearly polarized, 40~fs pulses.
	With increasing pump fluence of the exciting light field, (i)~the maximum of the absorption bleaches, (ii)~the absorption peak shifts towards higher energies, (iii)~the absorption broadens, and (iv)~for linearly polarized light a biexciton resonance appears.
	We identify the following mechanisms for the different observations:
	(i)~The decreasing nonlinear absorption for intensified pump fluences has two reasons -- firstly, Pauli blocking quenching the light-matter interaction, and secondly, exciton-exciton scattering redistributing the oscillator strength.
	However, in the weakly nonlinear regime the latter dominates and Pauli blocking is minor \cite{emmanuele2019highly}.
	(ii)~The excitation-induced shift of the excitonic resonance energy originates from Coulomb-induced band gap renormalization and mean field contributions.
	Whereas Hartree--Fock effects (second last contribution to Eq.~\eqref{eq:x-schematisch}) induce a blue shift, the exciton-exciton scattering continua (last contribution to Eq.~\eqref{eq:x-schematisch}) cause a smaller red shift.
	The interplay of both results in an effective blue shift \cite{blueshift} as observed in coherent pump-probe experiments \cite{mai2014exciton,sie2015valley,wang2015ultrafast,sie2017large,yong2018biexcitonic}.
	(iii)~The absorption spectra indicate a line broadening, cf. insets in Fig.~\ref{Bild-Exzitonen}~(a,b), associated with the formation of a pronounced shoulder on the high energy side of the exciton resonance.
	This nonlinear effect exclusively arises from the Coulomb coupling among excitons, biexcitons, and exciton-exciton scattering states.
	To analyze the dominating effects for the broadening, the four-particle density of states, defined as the imaginary part of the corresponding two-electron and two-hole scattering matrix in Supplementary Section~2, is plotted below the absorption spectra in Fig.~\ref{Bild-Exzitonen}~(a,b).
	The full four-particle density of states shows the exciton-exciton scattering continua above the exciton energy and an additional biexciton resonance below the exciton for linearly polarized excitation.
	This result clearly shows that nonlinear Coulomb scattering redistributes the oscillator strength and leads to pronounced shoulders on the high energy side of the exciton resonances.
	The effect resembles the non-Markovian coupling of excitons to the phonon continuum which also causes a shoulder on the high energy side accompanied by a polaron red shift \cite{christiansen2017phonon,shree2018observation,lengers2020theory}.
	However, whereas phonon sidebands disappear by lowering the temperature, the asymmetric lineshape due to exciton-exciton interactions remains.
	Thus, changing the temperature at elevated excitation power allows to distinguish both effects in optical experiments.
	(iv)~For excitation with a linearly polarized pulse the four-particle density of states exhibits a biexciton resulting in a resonance energetically below the exciton in Fig.~\ref{Bild-Exzitonen}~(b).
\textit{Analytical approach}:
	To provide an experimentally accessible evaluation of nonlinear line broadening effects in the coherent limit, we analytically extract the EID hereinafter.
	Applying a second-order Born--Markov approximation to eliminate the exciton-exciton scattering continua in Eq.~\eqref{eq:x-schematisch}, accesses the total linewidth of the energetically lowest exciton $P = P_{\text{1s}}$ in the $K$~valley which linearly increases with rising coherent exciton density~$\left|P\right|^2$:
	\begin{align}
	& \left[\hbar \partial_t + \gamma_0 - \text{i} \hat{\epsilon}_\text{x}  + \left(\gamma_\text{x} - \text{i} W_\text{x} \right) \left|P\right|^2 \right] P \notag \\
	& = -\text{i} \  {E}_{\sigma_+} \big({d} - \hat{{d}} \left|P\right|^2\big) . \label{eq-adel}
	\end{align}
	For the derivation, associated matrix elements, and comparison to the full numerical approach see Supplementary Section~3.
	Contrary to conventional III-V semiconductors, the EID in exciton dominated materials is determined by coherent exciton densities instead of incoherent electron-hole densities
	\cite{wang1993transient,hu1994excitation,wang1994transient}.
	Incoherent excitonic densities contribute only in the long time limit where exciton-phonon scattering dominates \cite{schmidt2016ultrafast,selig2017dark,selig2019ultrafast,christiansen2019theory,selig2019quenching}.
	Therefore, in the investigated case of (near) resonant excitation, EID expressed by $\gamma_\text{x}$ is determined by the exciton-exciton scattering continua solely.
	\begin{figure}
		\centering
		\begin{overpic}[width=0.95\columnwidth]{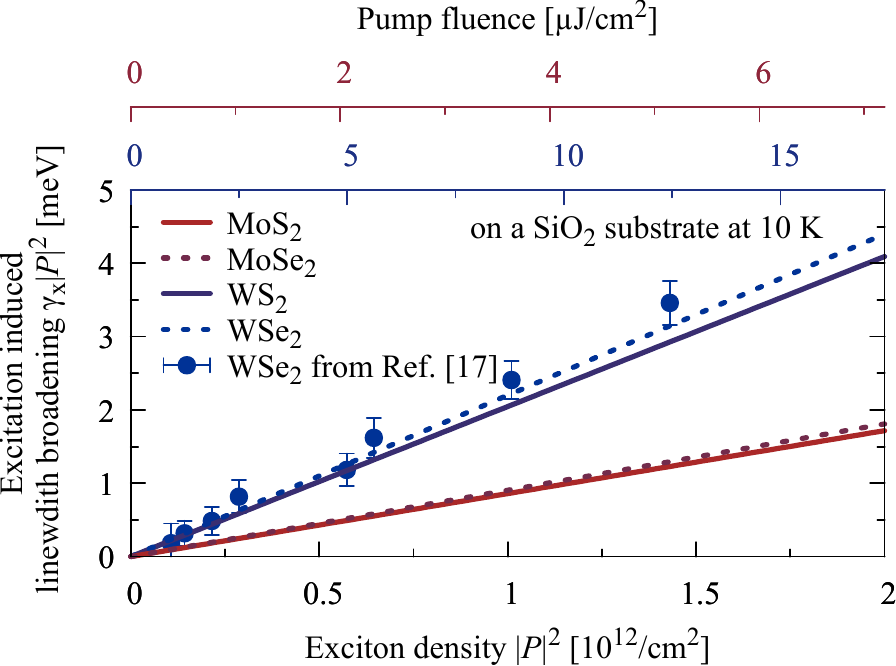}
			\put(0,71){(a)}
			\put(88,18){\large$\circlearrowright$}
		\end{overpic}\vspace{6mm}
	
		\begin{overpic}[width=0.95\columnwidth]{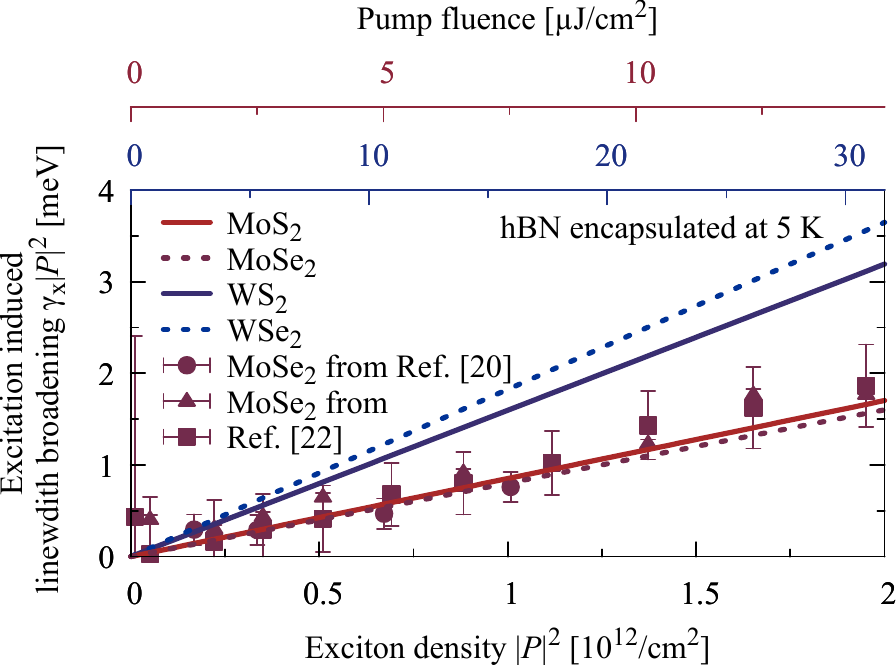}
			\put(0,71){(b)}
			\put(88,18){\large$\circlearrowright$}
		\end{overpic}
		\caption{\textbf{excitation-induced linewidth broadening.}
			(a)~Shown is the linearly increasing excitation-induced linewidth broadening $\gamma_\text{x}|P|^2$ depending on the exciton density $|P|^2$  for MoS$_2$, MoSe$_2$, WS$_2$, and WSe$_2$ on SiO$_2$ substrates at 10~K for circularly ($\circlearrowright$) polarized excitation.
			The pump fluence dependent linewidth increases are extracted from the absorption spectra obtained from numerically solving Eq.~\eqref{eq:x-schematisch} for 40~fs circularly polarized pulses.
			The dependence from the material absorption requires different axes roughly describing molybdenum (upper, red) and tungsten (lower, blue) based TMDCs.
			(b)~Same as in (a) but for TMDCs encapsulated in hBN at 5~K.
			The experimental data for WSe$_2$ and MoSe$_2$ adopted from Refs.~\cite{moody2015intrinsic,martin2018encapsulation,boule2020coherent} is divided by 1.1 and 0.3, respectively, to account for our calculations showing a 1.1 times larger or 0.3 times smaller excitation-induced linewidth increase measured in transmission and reflection compared to $\gamma_\text{x}|P|^2$, cf. Supplementary Section~4.
		}
		\label{Bild-EID}
	\end{figure}
	To show the impact of our calculations we analyze recent optical wave-mixing experiments \cite{moody2015intrinsic,martin2018encapsulation,boule2020coherent}:
	Fig.~\ref{Bild-EID}~(a) shows the microscopically calculated linear dependence between excitation-induced linewidth broadening~$\gamma_\text{x}\left|P\right|^2$ and exciton density assuming circularly polarized optical excitation for monolayer MoS$_2$, MoSe$_2$, WS$_2$, and WSe$_2$ on SiO$_2$ substrates at 10~K along with experimental data from Ref.~\cite{moody2015intrinsic}.
	WSe$_2$ exhibits the largest slope, closely followed by WS$_2$.
	For a direct experiment-theory comparison, the experimental detection geometry (transmission/reflection) has to be taken into account to compare the measured values to the calculated nonlinear susceptibility, cf. Supplementary Section~4.
	For the transmission geometry studied in Ref.~\cite{moody2015intrinsic}, we find a 1.1 times larger EID compared to the EID $\gamma_{\text{x}}$ of the nonlinear susceptibility, which requires to divide the experimental slope by 1.1.
	Our microscopically calculated EID for the nonlinear susceptibility $\gamma_\text{x} = 2.2 \cdot10^{-12}~\text{meV~cm}^2$ for WSe$_2$ is in very good agreement with the corrected measured value of $2.5 \cdot10^{-12}~\text{meV~cm}^2$ \cite{moody2015intrinsic}.
	Since molybdenum based TMDCs exhibit smaller conduction and valence band curvatures near the $K^{(\prime)}$~point compared with tungsten based TMDCs \cite{kormanyos2015k}, the EID for MoS$_2$ and MoSe$_2$ are a factor of two smaller.
	Fig.~\ref{Bild-EID}~(b) depicts the excitation-induced linewidth broadening for TMDCs encapsulated in hexagonal boron nitride (hBN) \cite{taniguchi2007synthesis} at 5~K.
	Here, the enhanced screening compared with SiO$_2$ substrates yields smaller EID.
	Our calculated EID $\gamma_\text{x} = 0.8 \cdot10^{-12}~\text{meV~cm}^2$ for MoSe$_2$ is in excellent agreement with recent experiments which measured a slope of $0.7 \cdot10^{-12}~\text{meV~cm}^2$ \cite{martin2018encapsulation} and $0.9 \cdot10^{-12}~\text{meV~cm}^2$ \cite{boule2020coherent}.
	To take the reflection geometry into account, where we predict a 0.3 times smaller EID compared to the nonlinear susceptibility $\gamma_{\text{x}}$, the experimental values had to be divided by 0.3 to compare them to our calculated values for the nonlinear susceptibility, cf. Supplementary Section~4.
	Assuming identical excitation conditions, we find that the largest values for the excitation-induced linewidth broadening of a material can be measured in absorption and the second largest values in transmission (both larger than $\gamma_\text{x}$).
	The smallest excitation-induced linewidth broadening is recorded in reflection (smaller than $\gamma_\text{x}$).
	
\textit{Conclusion}:
	We identified the coupling of excitons to exciton-exciton scattering continua as the dominating mechanism for the EID.
	The frequency-dependent linewidth broadening and shifts of the exciton resonance cause strong deviations from the symmetric Lorentzian lineshape.
	Our microscopical calculations are in good agreement with so far poorly understood experimental observations.
	We expect that our theoretical footing for EID is widely applicable to inorganic and organic semiconductors with tightly bound excitons indicating strong Coulomb interactions and pronounced many-body effects for nonlinear excitation below the Mott transition, or other bosonic gases in condensed matter physics \cite{afrousheh2004spectroscopic}.
	Possible candidates are not only other exciton dominated two-dimensional structures, such as back phosphorus \cite{liu2014phosphorene,tran2014layer,li2014black,rudenko2014quasiparticle,deilmann2018unraveling} or transition metal trichalcogenides \cite{dai2015titanium,van2018excitonic} but also, for instance, thin CdSe \cite{cassette2015room} or perovskite nanoplatelets \cite{wu2015excitonic,thouin2019enhanced}, and organic as well as hybrid organic-inorganic structures \cite{axt1998nonlinear,agranovich1998excitons,abdel2016exciton}.
	We believe that our work will trigger calculations in other material systems as our description is applicable once the excitonic eigenvalue problem is solved.

	\begin{acknowledgments}

		We thank Dominik Christiansen (TU Berlin) as well as Robert Schmidt and Rudolf Bratschitsch (WWU Münster) for stimulating discussions.
		Additionally, we thank Galan Moody (UC Santa Barbara), Elaine Li (UT Austin), Eric Martin and Steven Cundiff (University of Michigan), Jacek Kasprzak (CNRS Grenoble) and all involved coworkers for allowing us to use their experimental data.
		We gratefully acknowledge support from the Deutsche Forschungsgemeinschaft via the Projects No. 420760124 (KN 427/11-1, F.K. and A.K.) as well as No. 182087777---SFB 951 (B12, M.S. and A.K.).
		We also acknowledge the support of the European Unions Horizon 2020 research and innovation program under Grant Agreement No.~734690 (SONAR, A.K.).
		F.K. thanks the Berlin School of Optical Sciences and Quantum Technology.
	\end{acknowledgments}


\begin{thebibliography}{90}
	\expandafter\ifx\csname natexlab\endcsname\relax\def\natexlab#1{#1}\fi
	\expandafter\ifx\csname bibnamefont\endcsname\relax
	\def\bibnamefont#1{#1}\fi
	\expandafter\ifx\csname bibfnamefont\endcsname\relax
	\def\bibfnamefont#1{#1}\fi
	\expandafter\ifx\csname citenamefont\endcsname\relax
	\def\citenamefont#1{#1}\fi
	\expandafter\ifx\csname url\endcsname\relax
	\def\url#1{\texttt{#1}}\fi
	\expandafter\ifx\csname urlprefix\endcsname\relax\def\urlprefix{URL }\fi
	\providecommand{\bibinfo}[2]{#2}
	\providecommand{\eprint}[2][]{\url{#2}}
	
	\bibitem[{\citenamefont{Schultheis et~al.}(1986)\citenamefont{Schultheis, Kuhl,
			Honold, and Tu}}]{schultheis1986ultrafast}
	\bibinfo{author}{\bibfnamefont{L.}~\bibnamefont{Schultheis}},
	\bibinfo{author}{\bibfnamefont{J.}~\bibnamefont{Kuhl}},
	\bibinfo{author}{\bibfnamefont{A.}~\bibnamefont{Honold}}, \bibnamefont{and}
	\bibinfo{author}{\bibfnamefont{C.~W.} \bibnamefont{Tu}},
	\bibinfo{journal}{Physical Review Letters} \textbf{\bibinfo{volume}{57}},
	\bibinfo{pages}{1635} (\bibinfo{year}{1986}).
	
	\bibitem[{\citenamefont{Honold et~al.}(1989)\citenamefont{Honold, Schultheis,
			Kuhl, and Tu}}]{honold1989collision}
	\bibinfo{author}{\bibfnamefont{A.}~\bibnamefont{Honold}},
	\bibinfo{author}{\bibfnamefont{L.}~\bibnamefont{Schultheis}},
	\bibinfo{author}{\bibfnamefont{J.}~\bibnamefont{Kuhl}}, \bibnamefont{and}
	\bibinfo{author}{\bibfnamefont{C.~W.} \bibnamefont{Tu}},
	\bibinfo{journal}{Physical Review B} \textbf{\bibinfo{volume}{40}},
	\bibinfo{pages}{6442} (\bibinfo{year}{1989}).
	
	\bibitem[{\citenamefont{Wang et~al.}(1993)\citenamefont{Wang, Ferrio, Steel,
			Hu, Binder, and Koch}}]{wang1993transient}
	\bibinfo{author}{\bibfnamefont{H.}~\bibnamefont{Wang}},
	\bibinfo{author}{\bibfnamefont{K.}~\bibnamefont{Ferrio}},
	\bibinfo{author}{\bibfnamefont{D.~G.} \bibnamefont{Steel}},
	\bibinfo{author}{\bibfnamefont{Y.~Z.} \bibnamefont{Hu}},
	\bibinfo{author}{\bibfnamefont{R.}~\bibnamefont{Binder}}, \bibnamefont{and}
	\bibinfo{author}{\bibfnamefont{S.~W.} \bibnamefont{Koch}},
	\bibinfo{journal}{Physical Review Letters} \textbf{\bibinfo{volume}{71}},
	\bibinfo{pages}{1261} (\bibinfo{year}{1993}).
	
	\bibitem[{\citenamefont{Wang et~al.}(1994)\citenamefont{Wang, Ferrio, Steel,
			Berman, Hu, Binder, and Koch}}]{wang1994transient}
	\bibinfo{author}{\bibfnamefont{H.}~\bibnamefont{Wang}},
	\bibinfo{author}{\bibfnamefont{K.~B.} \bibnamefont{Ferrio}},
	\bibinfo{author}{\bibfnamefont{D.~G.} \bibnamefont{Steel}},
	\bibinfo{author}{\bibfnamefont{P.~R.} \bibnamefont{Berman}},
	\bibinfo{author}{\bibfnamefont{Y.~Z.} \bibnamefont{Hu}},
	\bibinfo{author}{\bibfnamefont{R.}~\bibnamefont{Binder}}, \bibnamefont{and}
	\bibinfo{author}{\bibfnamefont{S.~W.} \bibnamefont{Koch}},
	\bibinfo{journal}{Physical Review A} \textbf{\bibinfo{volume}{49}},
	\bibinfo{pages}{R1551} (\bibinfo{year}{1994}).
	
	\bibitem[{\citenamefont{Hu et~al.}(1994)\citenamefont{Hu, Binder, Koch,
			Cundiff, Wang, and Steel}}]{hu1994excitation}
	\bibinfo{author}{\bibfnamefont{Y.~Z.} \bibnamefont{Hu}},
	\bibinfo{author}{\bibfnamefont{R.}~\bibnamefont{Binder}},
	\bibinfo{author}{\bibfnamefont{S.~W.} \bibnamefont{Koch}},
	\bibinfo{author}{\bibfnamefont{S.~T.} \bibnamefont{Cundiff}},
	\bibinfo{author}{\bibfnamefont{H.}~\bibnamefont{Wang}}, \bibnamefont{and}
	\bibinfo{author}{\bibfnamefont{D.~G.} \bibnamefont{Steel}},
	\bibinfo{journal}{Physical Review B} \textbf{\bibinfo{volume}{49}},
	\bibinfo{pages}{14382} (\bibinfo{year}{1994}).
	
	\bibitem[{\citenamefont{Rappen et~al.}(1994)\citenamefont{Rappen, Peter,
			Wegener, and Sch{\"a}fer}}]{rappen1994polarization}
	\bibinfo{author}{\bibfnamefont{T.}~\bibnamefont{Rappen}},
	\bibinfo{author}{\bibfnamefont{U.-G.} \bibnamefont{Peter}},
	\bibinfo{author}{\bibfnamefont{M.}~\bibnamefont{Wegener}}, \bibnamefont{and}
	\bibinfo{author}{\bibfnamefont{W.}~\bibnamefont{Sch{\"a}fer}},
	\bibinfo{journal}{Physical Review B} \textbf{\bibinfo{volume}{49}},
	\bibinfo{pages}{10774} (\bibinfo{year}{1994}).
	
	\bibitem[{\citenamefont{Wagner et~al.}(1997)\citenamefont{Wagner, Sch{\"a}tz,
			Maier, Langbein, and Hvam}}]{wagner1997coherent}
	\bibinfo{author}{\bibfnamefont{H.~P.} \bibnamefont{Wagner}},
	\bibinfo{author}{\bibfnamefont{A.}~\bibnamefont{Sch{\"a}tz}},
	\bibinfo{author}{\bibfnamefont{R.}~\bibnamefont{Maier}},
	\bibinfo{author}{\bibfnamefont{W.}~\bibnamefont{Langbein}}, \bibnamefont{and}
	\bibinfo{author}{\bibfnamefont{J.~M.} \bibnamefont{Hvam}},
	\bibinfo{journal}{Physical Review B} \textbf{\bibinfo{volume}{56}},
	\bibinfo{pages}{12581} (\bibinfo{year}{1997}).
	
	\bibitem[{\citenamefont{Wagner et~al.}(1999)\citenamefont{Wagner, Sch{\"a}tz,
			Langbein, Hvam, and Smirl}}]{wagner1999interaction}
	\bibinfo{author}{\bibfnamefont{H.~P.} \bibnamefont{Wagner}},
	\bibinfo{author}{\bibfnamefont{A.}~\bibnamefont{Sch{\"a}tz}},
	\bibinfo{author}{\bibfnamefont{W.}~\bibnamefont{Langbein}},
	\bibinfo{author}{\bibfnamefont{J.~M.} \bibnamefont{Hvam}}, \bibnamefont{and}
	\bibinfo{author}{\bibfnamefont{A.~L.} \bibnamefont{Smirl}},
	\bibinfo{journal}{Physical Review B} \textbf{\bibinfo{volume}{60}},
	\bibinfo{pages}{4454} (\bibinfo{year}{1999}).
	
	\bibitem[{\citenamefont{Li et~al.}(2006)\citenamefont{Li, Zhang, Borca, and
			Cundiff}}]{li2006many}
	\bibinfo{author}{\bibfnamefont{X.}~\bibnamefont{Li}},
	\bibinfo{author}{\bibfnamefont{T.}~\bibnamefont{Zhang}},
	\bibinfo{author}{\bibfnamefont{C.~N.} \bibnamefont{Borca}}, \bibnamefont{and}
	\bibinfo{author}{\bibfnamefont{S.~T.} \bibnamefont{Cundiff}},
	\bibinfo{journal}{Physical Review Letters} \textbf{\bibinfo{volume}{96}},
	\bibinfo{pages}{057406} (\bibinfo{year}{2006}).
	
	\bibitem[{\citenamefont{Moody et~al.}(2011)\citenamefont{Moody, Siemens,
			Bristow, Dai, Karaiskaj, Bracker, Gammon, and Cundiff}}]{moody2011exciton}
	\bibinfo{author}{\bibfnamefont{G.}~\bibnamefont{Moody}},
	\bibinfo{author}{\bibfnamefont{M.~E.} \bibnamefont{Siemens}},
	\bibinfo{author}{\bibfnamefont{A.~D.} \bibnamefont{Bristow}},
	\bibinfo{author}{\bibfnamefont{X.}~\bibnamefont{Dai}},
	\bibinfo{author}{\bibfnamefont{D.}~\bibnamefont{Karaiskaj}},
	\bibinfo{author}{\bibfnamefont{A.~S.} \bibnamefont{Bracker}},
	\bibinfo{author}{\bibfnamefont{D.}~\bibnamefont{Gammon}}, \bibnamefont{and}
	\bibinfo{author}{\bibfnamefont{S.~T.} \bibnamefont{Cundiff}},
	\bibinfo{journal}{Physical Review B} \textbf{\bibinfo{volume}{83}},
	\bibinfo{pages}{115324} (\bibinfo{year}{2011}).
	
	\bibitem[{\citenamefont{Nardin et~al.}(2014)\citenamefont{Nardin, Moody, Singh,
			Autry, Li, Morier-Genoud, and Cundiff}}]{nardin2014coherent}
	\bibinfo{author}{\bibfnamefont{G.}~\bibnamefont{Nardin}},
	\bibinfo{author}{\bibfnamefont{G.}~\bibnamefont{Moody}},
	\bibinfo{author}{\bibfnamefont{R.}~\bibnamefont{Singh}},
	\bibinfo{author}{\bibfnamefont{T.~M.} \bibnamefont{Autry}},
	\bibinfo{author}{\bibfnamefont{H.}~\bibnamefont{Li}},
	\bibinfo{author}{\bibfnamefont{F.}~\bibnamefont{Morier-Genoud}},
	\bibnamefont{and} \bibinfo{author}{\bibfnamefont{S.~T.}
		\bibnamefont{Cundiff}}, \bibinfo{journal}{Physical Review Letters}
	\textbf{\bibinfo{volume}{112}}, \bibinfo{pages}{046402}
	(\bibinfo{year}{2014}).
	
	\bibitem[{\citenamefont{Novoselov et~al.}(2005)\citenamefont{Novoselov, Jiang,
			Schedin, Booth, Khotkevich, Morozov, and Geim}}]{novoselov2005two}
	\bibinfo{author}{\bibfnamefont{K.~S.} \bibnamefont{Novoselov}},
	\bibinfo{author}{\bibfnamefont{D.}~\bibnamefont{Jiang}},
	\bibinfo{author}{\bibfnamefont{F.}~\bibnamefont{Schedin}},
	\bibinfo{author}{\bibfnamefont{T.}~\bibnamefont{Booth}},
	\bibinfo{author}{\bibfnamefont{V.}~\bibnamefont{Khotkevich}},
	\bibinfo{author}{\bibfnamefont{S.}~\bibnamefont{Morozov}}, \bibnamefont{and}
	\bibinfo{author}{\bibfnamefont{A.~K.} \bibnamefont{Geim}},
	\bibinfo{journal}{Proceedings of the National Academy of Sciences}
	\textbf{\bibinfo{volume}{102}}, \bibinfo{pages}{10451}
	(\bibinfo{year}{2005}).
	
	\bibitem[{\citenamefont{Mak et~al.}(2010)\citenamefont{Mak, Lee, Hone, Shan,
			and Heinz}}]{mak2010atomically}
	\bibinfo{author}{\bibfnamefont{K.~F.} \bibnamefont{Mak}},
	\bibinfo{author}{\bibfnamefont{C.}~\bibnamefont{Lee}},
	\bibinfo{author}{\bibfnamefont{J.}~\bibnamefont{Hone}},
	\bibinfo{author}{\bibfnamefont{J.}~\bibnamefont{Shan}}, \bibnamefont{and}
	\bibinfo{author}{\bibfnamefont{T.~F.} \bibnamefont{Heinz}},
	\bibinfo{journal}{Physical Review Letters} \textbf{\bibinfo{volume}{105}},
	\bibinfo{pages}{136805} (\bibinfo{year}{2010}).
	
	\bibitem[{\citenamefont{Splendiani et~al.}(2010)\citenamefont{Splendiani, Sun,
			Zhang, Li, Kim, Chim, Galli, and Wang}}]{splendiani2010emerging}
	\bibinfo{author}{\bibfnamefont{A.}~\bibnamefont{Splendiani}},
	\bibinfo{author}{\bibfnamefont{L.}~\bibnamefont{Sun}},
	\bibinfo{author}{\bibfnamefont{Y.}~\bibnamefont{Zhang}},
	\bibinfo{author}{\bibfnamefont{T.}~\bibnamefont{Li}},
	\bibinfo{author}{\bibfnamefont{J.}~\bibnamefont{Kim}},
	\bibinfo{author}{\bibfnamefont{C.-Y.} \bibnamefont{Chim}},
	\bibinfo{author}{\bibfnamefont{G.}~\bibnamefont{Galli}}, \bibnamefont{and}
	\bibinfo{author}{\bibfnamefont{F.}~\bibnamefont{Wang}},
	\bibinfo{journal}{Nano Letters} \textbf{\bibinfo{volume}{10}},
	\bibinfo{pages}{1271} (\bibinfo{year}{2010}).
	
	\bibitem[{\citenamefont{Berkelbach et~al.}(2013)\citenamefont{Berkelbach,
			Hybertsen, and Reichman}}]{berkelbach2013theory}
	\bibinfo{author}{\bibfnamefont{T.~C.} \bibnamefont{Berkelbach}},
	\bibinfo{author}{\bibfnamefont{M.~S.} \bibnamefont{Hybertsen}},
	\bibnamefont{and} \bibinfo{author}{\bibfnamefont{D.~R.}
		\bibnamefont{Reichman}}, \bibinfo{journal}{Physical Review B}
	\textbf{\bibinfo{volume}{88}}, \bibinfo{pages}{045318}
	(\bibinfo{year}{2013}).
	
	\bibitem[{\citenamefont{Chernikov et~al.}(2014)\citenamefont{Chernikov,
			Berkelbach, Hill, Rigosi, Li, Aslan, Reichman, Hybertsen, and
			Heinz}}]{chernikov2014exciton}
	\bibinfo{author}{\bibfnamefont{A.}~\bibnamefont{Chernikov}},
	\bibinfo{author}{\bibfnamefont{T.~C.} \bibnamefont{Berkelbach}},
	\bibinfo{author}{\bibfnamefont{H.~M.} \bibnamefont{Hill}},
	\bibinfo{author}{\bibfnamefont{A.}~\bibnamefont{Rigosi}},
	\bibinfo{author}{\bibfnamefont{Y.}~\bibnamefont{Li}},
	\bibinfo{author}{\bibfnamefont{O.~B.} \bibnamefont{Aslan}},
	\bibinfo{author}{\bibfnamefont{D.~R.} \bibnamefont{Reichman}},
	\bibinfo{author}{\bibfnamefont{M.~S.} \bibnamefont{Hybertsen}},
	\bibnamefont{and} \bibinfo{author}{\bibfnamefont{T.~F.} \bibnamefont{Heinz}},
	\bibinfo{journal}{Physical Review Letters} \textbf{\bibinfo{volume}{113}},
	\bibinfo{pages}{076802} (\bibinfo{year}{2014}).
	
	\bibitem[{\citenamefont{Moody et~al.}(2015)\citenamefont{Moody, Dass, Hao,
			Chen, Li, Singh, Tran, Clark, Xu, Bergh{\"a}user
			et~al.}}]{moody2015intrinsic}
	\bibinfo{author}{\bibfnamefont{G.}~\bibnamefont{Moody}},
	\bibinfo{author}{\bibfnamefont{C.~K.} \bibnamefont{Dass}},
	\bibinfo{author}{\bibfnamefont{K.}~\bibnamefont{Hao}},
	\bibinfo{author}{\bibfnamefont{C.-H.} \bibnamefont{Chen}},
	\bibinfo{author}{\bibfnamefont{L.-J.} \bibnamefont{Li}},
	\bibinfo{author}{\bibfnamefont{A.}~\bibnamefont{Singh}},
	\bibinfo{author}{\bibfnamefont{K.}~\bibnamefont{Tran}},
	\bibinfo{author}{\bibfnamefont{G.}~\bibnamefont{Clark}},
	\bibinfo{author}{\bibfnamefont{X.}~\bibnamefont{Xu}},
	\bibinfo{author}{\bibfnamefont{G.}~\bibnamefont{Bergh{\"a}user}},
	\bibnamefont{et~al.}, \bibinfo{journal}{Nature Communications}
	\textbf{\bibinfo{volume}{6}}, \bibinfo{pages}{8315} (\bibinfo{year}{2015}).
	
	\bibitem[{\citenamefont{Dey et~al.}(2016)\citenamefont{Dey, Paul, Wang,
			Stevens, Liu, Romero, Shan, Hilton, and Karaiskaj}}]{dey2016optical}
	\bibinfo{author}{\bibfnamefont{P.}~\bibnamefont{Dey}},
	\bibinfo{author}{\bibfnamefont{J.}~\bibnamefont{Paul}},
	\bibinfo{author}{\bibfnamefont{Z.}~\bibnamefont{Wang}},
	\bibinfo{author}{\bibfnamefont{C.~E.} \bibnamefont{Stevens}},
	\bibinfo{author}{\bibfnamefont{C.}~\bibnamefont{Liu}},
	\bibinfo{author}{\bibfnamefont{A.~H.} \bibnamefont{Romero}},
	\bibinfo{author}{\bibfnamefont{J.}~\bibnamefont{Shan}},
	\bibinfo{author}{\bibfnamefont{D.~J.} \bibnamefont{Hilton}},
	\bibnamefont{and}
	\bibinfo{author}{\bibfnamefont{D.}~\bibnamefont{Karaiskaj}},
	\bibinfo{journal}{Physical Review Letters} \textbf{\bibinfo{volume}{116}},
	\bibinfo{pages}{127402} (\bibinfo{year}{2016}).
	
	\bibitem[{\citenamefont{Mahmood et~al.}(2017)\citenamefont{Mahmood, Alpichshev,
			Lee, Kong, and Gedik}}]{mahmood2017observation}
	\bibinfo{author}{\bibfnamefont{F.}~\bibnamefont{Mahmood}},
	\bibinfo{author}{\bibfnamefont{Z.}~\bibnamefont{Alpichshev}},
	\bibinfo{author}{\bibfnamefont{Y.-H.} \bibnamefont{Lee}},
	\bibinfo{author}{\bibfnamefont{J.}~\bibnamefont{Kong}}, \bibnamefont{and}
	\bibinfo{author}{\bibfnamefont{N.}~\bibnamefont{Gedik}},
	\bibinfo{journal}{Nano Letters} \textbf{\bibinfo{volume}{18}},
	\bibinfo{pages}{223} (\bibinfo{year}{2017}).
	
	\bibitem[{\citenamefont{Martin et~al.}(2018)\citenamefont{Martin, Horng, Ruth,
			Paik, Wentzel, Deng, and Cundiff}}]{martin2018encapsulation}
	\bibinfo{author}{\bibfnamefont{E.~W.} \bibnamefont{Martin}},
	\bibinfo{author}{\bibfnamefont{J.}~\bibnamefont{Horng}},
	\bibinfo{author}{\bibfnamefont{H.~G.} \bibnamefont{Ruth}},
	\bibinfo{author}{\bibfnamefont{E.}~\bibnamefont{Paik}},
	\bibinfo{author}{\bibfnamefont{M.-H.} \bibnamefont{Wentzel}},
	\bibinfo{author}{\bibfnamefont{H.}~\bibnamefont{Deng}}, \bibnamefont{and}
	\bibinfo{author}{\bibfnamefont{S.~T.} \bibnamefont{Cundiff}},
	\bibinfo{journal}{arXiv preprint arXiv:1810.09834}  (\bibinfo{year}{2018}).
	
	\bibitem[{\citenamefont{Jakubczyk et~al.}(2019)\citenamefont{Jakubczyk, Nayak,
			Scarpelli, Liu, Dubey, Bendiab, Marty, Taniguchi, Watanabe, Masia
			et~al.}}]{jakubczyk2019coherence}
	\bibinfo{author}{\bibfnamefont{T.}~\bibnamefont{Jakubczyk}},
	\bibinfo{author}{\bibfnamefont{G.}~\bibnamefont{Nayak}},
	\bibinfo{author}{\bibfnamefont{L.}~\bibnamefont{Scarpelli}},
	\bibinfo{author}{\bibfnamefont{W.-L.} \bibnamefont{Liu}},
	\bibinfo{author}{\bibfnamefont{S.}~\bibnamefont{Dubey}},
	\bibinfo{author}{\bibfnamefont{N.}~\bibnamefont{Bendiab}},
	\bibinfo{author}{\bibfnamefont{L.}~\bibnamefont{Marty}},
	\bibinfo{author}{\bibfnamefont{T.}~\bibnamefont{Taniguchi}},
	\bibinfo{author}{\bibfnamefont{K.}~\bibnamefont{Watanabe}},
	\bibinfo{author}{\bibfnamefont{F.}~\bibnamefont{Masia}},
	\bibnamefont{et~al.}, \bibinfo{journal}{ACS Nano}
	\textbf{\bibinfo{volume}{13}}, \bibinfo{pages}{3500} (\bibinfo{year}{2019}).
	
	\bibitem[{\citenamefont{Boule et~al.}(2020)\citenamefont{Boule, Vaclavkova,
			Bartos, Nogajewski, Zdra{\v{z}}il, Taniguchi, Watanabe, Potemski, and
			Kasprzak}}]{boule2020coherent}
	\bibinfo{author}{\bibfnamefont{C.}~\bibnamefont{Boule}},
	\bibinfo{author}{\bibfnamefont{D.}~\bibnamefont{Vaclavkova}},
	\bibinfo{author}{\bibfnamefont{M.}~\bibnamefont{Bartos}},
	\bibinfo{author}{\bibfnamefont{K.}~\bibnamefont{Nogajewski}},
	\bibinfo{author}{\bibfnamefont{L.}~\bibnamefont{Zdra{\v{z}}il}},
	\bibinfo{author}{\bibfnamefont{T.}~\bibnamefont{Taniguchi}},
	\bibinfo{author}{\bibfnamefont{K.}~\bibnamefont{Watanabe}},
	\bibinfo{author}{\bibfnamefont{M.}~\bibnamefont{Potemski}}, \bibnamefont{and}
	\bibinfo{author}{\bibfnamefont{J.}~\bibnamefont{Kasprzak}},
	\bibinfo{journal}{Physical Review Materials} \textbf{\bibinfo{volume}{4}},
	\bibinfo{pages}{034001} (\bibinfo{year}{2020}).
	
	\bibitem[{\citenamefont{Ivanov and Haug}(1993)}]{ivanov1993self}
	\bibinfo{author}{\bibfnamefont{A.~L.} \bibnamefont{Ivanov}} \bibnamefont{and}
	\bibinfo{author}{\bibfnamefont{H.}~\bibnamefont{Haug}},
	\bibinfo{journal}{Physical Review B} \textbf{\bibinfo{volume}{48}},
	\bibinfo{pages}{1490} (\bibinfo{year}{1993}).
	
	\bibitem[{\citenamefont{Katsch et~al.}(2018)\citenamefont{Katsch, Selig,
			Carmele, and Knorr}}]{katsch2018theory}
	\bibinfo{author}{\bibfnamefont{F.}~\bibnamefont{Katsch}},
	\bibinfo{author}{\bibfnamefont{M.}~\bibnamefont{Selig}},
	\bibinfo{author}{\bibfnamefont{A.}~\bibnamefont{Carmele}}, \bibnamefont{and}
	\bibinfo{author}{\bibfnamefont{A.}~\bibnamefont{Knorr}},
	\bibinfo{journal}{Physica Status Solidi (b)} \textbf{\bibinfo{volume}{255}},
	\bibinfo{pages}{1800185} (\bibinfo{year}{2018}).
	
	\bibitem[{\citenamefont{Kyl{\"a}np{\"a}{\"a} and
			Komsa}(2015)}]{kylanpaa2015binding}
	\bibinfo{author}{\bibfnamefont{I.}~\bibnamefont{Kyl{\"a}np{\"a}{\"a}}}
	\bibnamefont{and} \bibinfo{author}{\bibfnamefont{H.-P.} \bibnamefont{Komsa}},
	\bibinfo{journal}{Physical Review B} \textbf{\bibinfo{volume}{92}},
	\bibinfo{pages}{205418} (\bibinfo{year}{2015}).
	
	\bibitem[{\citenamefont{Mayers et~al.}(2015)\citenamefont{Mayers, Berkelbach,
			Hybertsen, and Reichman}}]{mayers2015binding}
	\bibinfo{author}{\bibfnamefont{M.~Z.} \bibnamefont{Mayers}},
	\bibinfo{author}{\bibfnamefont{T.~C.} \bibnamefont{Berkelbach}},
	\bibinfo{author}{\bibfnamefont{M.~S.} \bibnamefont{Hybertsen}},
	\bibnamefont{and} \bibinfo{author}{\bibfnamefont{D.~R.}
		\bibnamefont{Reichman}}, \bibinfo{journal}{Physical Review B}
	\textbf{\bibinfo{volume}{92}}, \bibinfo{pages}{161404(R)}
	(\bibinfo{year}{2015}).
	
	\bibitem[{\citenamefont{Zhang et~al.}(2015)\citenamefont{Zhang, Kidd, and
			Varga}}]{zhang2015excited}
	\bibinfo{author}{\bibfnamefont{D.~K.} \bibnamefont{Zhang}},
	\bibinfo{author}{\bibfnamefont{D.~W.} \bibnamefont{Kidd}}, \bibnamefont{and}
	\bibinfo{author}{\bibfnamefont{K.}~\bibnamefont{Varga}},
	\bibinfo{journal}{Nano Letters} \textbf{\bibinfo{volume}{15}},
	\bibinfo{pages}{7002} (\bibinfo{year}{2015}).
	
	\bibitem[{\citenamefont{Kidd et~al.}(2016)\citenamefont{Kidd, Zhang, and
			Varga}}]{kidd2016binding}
	\bibinfo{author}{\bibfnamefont{D.~W.} \bibnamefont{Kidd}},
	\bibinfo{author}{\bibfnamefont{D.~K.} \bibnamefont{Zhang}}, \bibnamefont{and}
	\bibinfo{author}{\bibfnamefont{K.}~\bibnamefont{Varga}},
	\bibinfo{journal}{Physical Review B} \textbf{\bibinfo{volume}{93}},
	\bibinfo{pages}{125423} (\bibinfo{year}{2016}).
	
	\bibitem[{\citenamefont{Mostaani et~al.}(2017)\citenamefont{Mostaani,
			Szyniszewski, Price, Maezono, Danovich, Hunt, Drummond, and
			Fal'ko}}]{mostaani2017diffusion}
	\bibinfo{author}{\bibfnamefont{E.}~\bibnamefont{Mostaani}},
	\bibinfo{author}{\bibfnamefont{M.}~\bibnamefont{Szyniszewski}},
	\bibinfo{author}{\bibfnamefont{C.~H.} \bibnamefont{Price}},
	\bibinfo{author}{\bibfnamefont{R.}~\bibnamefont{Maezono}},
	\bibinfo{author}{\bibfnamefont{M.}~\bibnamefont{Danovich}},
	\bibinfo{author}{\bibfnamefont{R.~J.} \bibnamefont{Hunt}},
	\bibinfo{author}{\bibfnamefont{N.~D.} \bibnamefont{Drummond}},
	\bibnamefont{and} \bibinfo{author}{\bibfnamefont{V.~I.}
		\bibnamefont{Fal'ko}}, \bibinfo{journal}{Physical Review B}
	\textbf{\bibinfo{volume}{96}}, \bibinfo{pages}{075431}
	(\bibinfo{year}{2017}).
	
	\bibitem[{\citenamefont{Kezerashvili and
			Tsiklauri}(2017)}]{kezerashvili2017trion}
	\bibinfo{author}{\bibfnamefont{R.~Y.} \bibnamefont{Kezerashvili}}
	\bibnamefont{and} \bibinfo{author}{\bibfnamefont{S.~M.}
		\bibnamefont{Tsiklauri}}, \bibinfo{journal}{Few-Body Systems}
	\textbf{\bibinfo{volume}{58}}, \bibinfo{pages}{18} (\bibinfo{year}{2017}).
	
	\bibitem[{\citenamefont{Szyniszewski et~al.}(2017)\citenamefont{Szyniszewski,
			Mostaani, Drummond, and Fal'ko}}]{szyniszewski2017binding}
	\bibinfo{author}{\bibfnamefont{M.}~\bibnamefont{Szyniszewski}},
	\bibinfo{author}{\bibfnamefont{E.}~\bibnamefont{Mostaani}},
	\bibinfo{author}{\bibfnamefont{N.~D.} \bibnamefont{Drummond}},
	\bibnamefont{and} \bibinfo{author}{\bibfnamefont{V.~I.}
		\bibnamefont{Fal'ko}}, \bibinfo{journal}{Physical Review B}
	\textbf{\bibinfo{volume}{95}}, \bibinfo{pages}{081301(R)}
	(\bibinfo{year}{2017}).
	
	\bibitem[{\citenamefont{Van~der Donck et~al.}(2018)\citenamefont{Van~der Donck,
			Zarenia, and Peeters}}]{van2018excitons}
	\bibinfo{author}{\bibfnamefont{M.}~\bibnamefont{Van~der Donck}},
	\bibinfo{author}{\bibfnamefont{M.}~\bibnamefont{Zarenia}}, \bibnamefont{and}
	\bibinfo{author}{\bibfnamefont{F.~M.} \bibnamefont{Peeters}},
	\bibinfo{journal}{Physical Review B} \textbf{\bibinfo{volume}{97}},
	\bibinfo{pages}{195408} (\bibinfo{year}{2018}).
	
	\bibitem[{\citenamefont{Steinhoff et~al.}(2018)\citenamefont{Steinhoff,
			Florian, Singh, Tran, Kolarczik, Helmrich, Achtstein, Woggon, Owschimikow,
			Jahnke et~al.}}]{steinhoff2018biexciton}
	\bibinfo{author}{\bibfnamefont{A.}~\bibnamefont{Steinhoff}},
	\bibinfo{author}{\bibfnamefont{M.}~\bibnamefont{Florian}},
	\bibinfo{author}{\bibfnamefont{A.}~\bibnamefont{Singh}},
	\bibinfo{author}{\bibfnamefont{K.}~\bibnamefont{Tran}},
	\bibinfo{author}{\bibfnamefont{M.}~\bibnamefont{Kolarczik}},
	\bibinfo{author}{\bibfnamefont{S.}~\bibnamefont{Helmrich}},
	\bibinfo{author}{\bibfnamefont{A.~W.} \bibnamefont{Achtstein}},
	\bibinfo{author}{\bibfnamefont{U.}~\bibnamefont{Woggon}},
	\bibinfo{author}{\bibfnamefont{N.}~\bibnamefont{Owschimikow}},
	\bibinfo{author}{\bibfnamefont{F.}~\bibnamefont{Jahnke}},
	\bibnamefont{et~al.}, \bibinfo{journal}{Nature Physics}
	\textbf{\bibinfo{volume}{14}}, \bibinfo{pages}{1199} (\bibinfo{year}{2018}).
	
	\bibitem[{\citenamefont{Kuhn and Richter}(2019)}]{kuhn2019combined}
	\bibinfo{author}{\bibfnamefont{S.~C.} \bibnamefont{Kuhn}} \bibnamefont{and}
	\bibinfo{author}{\bibfnamefont{M.}~\bibnamefont{Richter}},
	\bibinfo{journal}{Physical Review B} \textbf{\bibinfo{volume}{99}},
	\bibinfo{pages}{241301(R)} (\bibinfo{year}{2019}).
	
	\bibitem[{\citenamefont{Sie et~al.}(2015{\natexlab{a}})\citenamefont{Sie,
			Frenzel, Lee, Kong, and Gedik}}]{sie2015intervalley}
	\bibinfo{author}{\bibfnamefont{E.~J.} \bibnamefont{Sie}},
	\bibinfo{author}{\bibfnamefont{A.~J.} \bibnamefont{Frenzel}},
	\bibinfo{author}{\bibfnamefont{Y.-H.} \bibnamefont{Lee}},
	\bibinfo{author}{\bibfnamefont{J.}~\bibnamefont{Kong}}, \bibnamefont{and}
	\bibinfo{author}{\bibfnamefont{N.}~\bibnamefont{Gedik}},
	\bibinfo{journal}{Physical Revkiew B} \textbf{\bibinfo{volume}{92}},
	\bibinfo{pages}{125417} (\bibinfo{year}{2015}{\natexlab{a}}).
	
	\bibitem[{\citenamefont{Aleithan et~al.}(2016)\citenamefont{Aleithan, Livshits,
			Khadka, Rack, Kordesch, and Stinaff}}]{aleithan2016broadband}
	\bibinfo{author}{\bibfnamefont{S.~H.} \bibnamefont{Aleithan}},
	\bibinfo{author}{\bibfnamefont{M.~Y.} \bibnamefont{Livshits}},
	\bibinfo{author}{\bibfnamefont{S.}~\bibnamefont{Khadka}},
	\bibinfo{author}{\bibfnamefont{J.~J.} \bibnamefont{Rack}},
	\bibinfo{author}{\bibfnamefont{M.~E.} \bibnamefont{Kordesch}},
	\bibnamefont{and} \bibinfo{author}{\bibfnamefont{E.}~\bibnamefont{Stinaff}},
	\bibinfo{journal}{Physical Review B} \textbf{\bibinfo{volume}{94}},
	\bibinfo{pages}{035445} (\bibinfo{year}{2016}).
	
	\bibitem[{\citenamefont{Sie et~al.}(2016)\citenamefont{Sie, Lui, Lee, Kong, and
			Gedik}}]{sie2016observation}
	\bibinfo{author}{\bibfnamefont{E.~J.} \bibnamefont{Sie}},
	\bibinfo{author}{\bibfnamefont{C.~H.} \bibnamefont{Lui}},
	\bibinfo{author}{\bibfnamefont{Y.-H.} \bibnamefont{Lee}},
	\bibinfo{author}{\bibfnamefont{J.}~\bibnamefont{Kong}}, \bibnamefont{and}
	\bibinfo{author}{\bibfnamefont{N.}~\bibnamefont{Gedik}},
	\bibinfo{journal}{Nano Letters} \textbf{\bibinfo{volume}{16}},
	\bibinfo{pages}{7421} (\bibinfo{year}{2016}).
	
	\bibitem[{\citenamefont{Hao et~al.}(2017)\citenamefont{Hao, Specht, Nagler, Xu,
			Tran, Singh, Dass, Sch{\"u}ller, Korn, Richter et~al.}}]{hao2017neutral}
	\bibinfo{author}{\bibfnamefont{K.}~\bibnamefont{Hao}},
	\bibinfo{author}{\bibfnamefont{J.~F.} \bibnamefont{Specht}},
	\bibinfo{author}{\bibfnamefont{P.}~\bibnamefont{Nagler}},
	\bibinfo{author}{\bibfnamefont{L.}~\bibnamefont{Xu}},
	\bibinfo{author}{\bibfnamefont{K.}~\bibnamefont{Tran}},
	\bibinfo{author}{\bibfnamefont{A.}~\bibnamefont{Singh}},
	\bibinfo{author}{\bibfnamefont{C.~K.} \bibnamefont{Dass}},
	\bibinfo{author}{\bibfnamefont{C.}~\bibnamefont{Sch{\"u}ller}},
	\bibinfo{author}{\bibfnamefont{T.}~\bibnamefont{Korn}},
	\bibinfo{author}{\bibfnamefont{M.}~\bibnamefont{Richter}},
	\bibnamefont{et~al.}, \bibinfo{journal}{Nature Communications}
	\textbf{\bibinfo{volume}{8}} (\bibinfo{year}{2017}).
	
	\bibitem[{\citenamefont{Axt et~al.}(1998)\citenamefont{Axt, Victor, and
			Kuhn}}]{axt1998exciton}
	\bibinfo{author}{\bibfnamefont{V.}~\bibnamefont{Axt}},
	\bibinfo{author}{\bibfnamefont{K.}~\bibnamefont{Victor}}, \bibnamefont{and}
	\bibinfo{author}{\bibfnamefont{T.}~\bibnamefont{Kuhn}},
	\bibinfo{journal}{Physica Status Solidi (b)} \textbf{\bibinfo{volume}{206}},
	\bibinfo{pages}{189} (\bibinfo{year}{1998}).
	
	\bibitem[{\citenamefont{Bartels et~al.}(1998)\citenamefont{Bartels, Stahl, Axt,
			Haase, Neukirch, and Gutowski}}]{bartels1998identification}
	\bibinfo{author}{\bibfnamefont{G.}~\bibnamefont{Bartels}},
	\bibinfo{author}{\bibfnamefont{A.}~\bibnamefont{Stahl}},
	\bibinfo{author}{\bibfnamefont{V.~M.} \bibnamefont{Axt}},
	\bibinfo{author}{\bibfnamefont{B.}~\bibnamefont{Haase}},
	\bibinfo{author}{\bibfnamefont{U.}~\bibnamefont{Neukirch}}, \bibnamefont{and}
	\bibinfo{author}{\bibfnamefont{J.}~\bibnamefont{Gutowski}},
	\bibinfo{journal}{Physical Review Letters} \textbf{\bibinfo{volume}{81}},
	\bibinfo{pages}{5880} (\bibinfo{year}{1998}).
	
	\bibitem[{\citenamefont{Kwong et~al.}(2001{\natexlab{a}})\citenamefont{Kwong,
			Takayama, Rumyantsev, Kuwata-Gonokami, and Binder}}]{kwong2001third}
	\bibinfo{author}{\bibfnamefont{N.~H.} \bibnamefont{Kwong}},
	\bibinfo{author}{\bibfnamefont{R.}~\bibnamefont{Takayama}},
	\bibinfo{author}{\bibfnamefont{I.}~\bibnamefont{Rumyantsev}},
	\bibinfo{author}{\bibfnamefont{M.}~\bibnamefont{Kuwata-Gonokami}},
	\bibnamefont{and} \bibinfo{author}{\bibfnamefont{R.}~\bibnamefont{Binder}},
	\bibinfo{journal}{Physical Review B} \textbf{\bibinfo{volume}{64}},
	\bibinfo{pages}{045316} (\bibinfo{year}{2001}{\natexlab{a}}).
	
	\bibitem[{\citenamefont{Kwong et~al.}(2001{\natexlab{b}})\citenamefont{Kwong,
			Takayama, Rumyantsev, Kuwata-Gonokami, and Binder}}]{kwong2001evidence}
	\bibinfo{author}{\bibfnamefont{N.-H.} \bibnamefont{Kwong}},
	\bibinfo{author}{\bibfnamefont{R.}~\bibnamefont{Takayama}},
	\bibinfo{author}{\bibfnamefont{I.}~\bibnamefont{Rumyantsev}},
	\bibinfo{author}{\bibfnamefont{M.}~\bibnamefont{Kuwata-Gonokami}},
	\bibnamefont{and} \bibinfo{author}{\bibfnamefont{R.}~\bibnamefont{Binder}},
	\bibinfo{journal}{Physical Review Letters} \textbf{\bibinfo{volume}{87}},
	\bibinfo{pages}{027402} (\bibinfo{year}{2001}{\natexlab{b}}).
	
	\bibitem[{\citenamefont{Takayama et~al.}(2002)\citenamefont{Takayama, Kwong,
			Rumyantsev, Kuwata-Gonokami, and Binder}}]{takayama2002t}
	\bibinfo{author}{\bibfnamefont{R.}~\bibnamefont{Takayama}},
	\bibinfo{author}{\bibfnamefont{N.}~\bibnamefont{Kwong}},
	\bibinfo{author}{\bibfnamefont{I.}~\bibnamefont{Rumyantsev}},
	\bibinfo{author}{\bibfnamefont{M.}~\bibnamefont{Kuwata-Gonokami}},
	\bibnamefont{and} \bibinfo{author}{\bibfnamefont{R.}~\bibnamefont{Binder}},
	\bibinfo{journal}{The European Physical Journal B-Condensed Matter and
		Complex Systems} \textbf{\bibinfo{volume}{25}}, \bibinfo{pages}{445}
	(\bibinfo{year}{2002}).
	
	\bibitem[{\citenamefont{Voss et~al.}(2002)\citenamefont{Voss, Breunig,
			R{\"u}ckmann, Gutowski, Axt, and Kuhn}}]{voss2002biexcitonic}
	\bibinfo{author}{\bibfnamefont{T.}~\bibnamefont{Voss}},
	\bibinfo{author}{\bibfnamefont{H.~G.} \bibnamefont{Breunig}},
	\bibinfo{author}{\bibfnamefont{I.}~\bibnamefont{R{\"u}ckmann}},
	\bibinfo{author}{\bibfnamefont{J.}~\bibnamefont{Gutowski}},
	\bibinfo{author}{\bibfnamefont{V.~M.} \bibnamefont{Axt}}, \bibnamefont{and}
	\bibinfo{author}{\bibfnamefont{T.}~\bibnamefont{Kuhn}},
	\bibinfo{journal}{Physical Review B} \textbf{\bibinfo{volume}{66}},
	\bibinfo{pages}{155301} (\bibinfo{year}{2002}).
	
	\bibitem[{\citenamefont{Schumacher et~al.}(2005)\citenamefont{Schumacher,
			Czycholl, Jahnke, Kudyk, Wischmeier, R{\"u}ckmann, Voss, Gutowski, Gust, and
			Hommel}}]{schumacher2005coherent}
	\bibinfo{author}{\bibfnamefont{S.}~\bibnamefont{Schumacher}},
	\bibinfo{author}{\bibfnamefont{G.}~\bibnamefont{Czycholl}},
	\bibinfo{author}{\bibfnamefont{F.}~\bibnamefont{Jahnke}},
	\bibinfo{author}{\bibfnamefont{I.}~\bibnamefont{Kudyk}},
	\bibinfo{author}{\bibfnamefont{L.}~\bibnamefont{Wischmeier}},
	\bibinfo{author}{\bibfnamefont{I.}~\bibnamefont{R{\"u}ckmann}},
	\bibinfo{author}{\bibfnamefont{T.}~\bibnamefont{Voss}},
	\bibinfo{author}{\bibfnamefont{J.}~\bibnamefont{Gutowski}},
	\bibinfo{author}{\bibfnamefont{A.}~\bibnamefont{Gust}}, \bibnamefont{and}
	\bibinfo{author}{\bibfnamefont{D.}~\bibnamefont{Hommel}},
	\bibinfo{journal}{Physical Review B} \textbf{\bibinfo{volume}{72}},
	\bibinfo{pages}{081308(R)} (\bibinfo{year}{2005}).
	
	\bibitem[{\citenamefont{Schumacher et~al.}(2006)\citenamefont{Schumacher,
			Czycholl, and Jahnke}}]{schumacher2006coherent}
	\bibinfo{author}{\bibfnamefont{S.}~\bibnamefont{Schumacher}},
	\bibinfo{author}{\bibfnamefont{G.}~\bibnamefont{Czycholl}}, \bibnamefont{and}
	\bibinfo{author}{\bibfnamefont{F.}~\bibnamefont{Jahnke}},
	\bibinfo{journal}{Physical Review B} \textbf{\bibinfo{volume}{73}},
	\bibinfo{pages}{035318} (\bibinfo{year}{2006}).
	
	\bibitem[{\citenamefont{Voss et~al.}(2006)\citenamefont{Voss, R{\"u}ckmann,
			Gutowski, Axt, and Kuhn}}]{voss2006coherent}
	\bibinfo{author}{\bibfnamefont{T.}~\bibnamefont{Voss}},
	\bibinfo{author}{\bibfnamefont{I.}~\bibnamefont{R{\"u}ckmann}},
	\bibinfo{author}{\bibfnamefont{J.}~\bibnamefont{Gutowski}},
	\bibinfo{author}{\bibfnamefont{V.~M.} \bibnamefont{Axt}}, \bibnamefont{and}
	\bibinfo{author}{\bibfnamefont{T.}~\bibnamefont{Kuhn}},
	\bibinfo{journal}{Physical Review B} \textbf{\bibinfo{volume}{73}},
	\bibinfo{pages}{115311} (\bibinfo{year}{2006}).
	
	\bibitem[{\citenamefont{Schumacher et~al.}(2007)\citenamefont{Schumacher,
			Kwong, and Binder}}]{schumacher2007influence}
	\bibinfo{author}{\bibfnamefont{S.}~\bibnamefont{Schumacher}},
	\bibinfo{author}{\bibfnamefont{N.-H.} \bibnamefont{Kwong}}, \bibnamefont{and}
	\bibinfo{author}{\bibfnamefont{R.}~\bibnamefont{Binder}},
	\bibinfo{journal}{Physical Review B} \textbf{\bibinfo{volume}{76}},
	\bibinfo{pages}{245324} (\bibinfo{year}{2007}).
	
	\bibitem[{\citenamefont{Sch{\"a}fer and
			Wege\textbf{}ner}(2013)}]{schafer2013semiconductor}
	\bibinfo{author}{\bibfnamefont{W.}~\bibnamefont{Sch{\"a}fer}} \bibnamefont{and}
	\bibinfo{author}{\bibfnamefont{M.}~\bibnamefont{Wege\textbf{}ner}},
	\emph{\bibinfo{title}{{Semiconductor optics and transport phenomena}}}
	(\bibinfo{publisher}{Springer Science \& Business Media},
	\bibinfo{year}{2013}).
	
	\bibitem[{\citenamefont{Katsch et~al.}(2020)\citenamefont{Katsch, Selig, and
			Knorr}}]{katsch2020theory}
	\bibinfo{author}{\bibfnamefont{F.}~\bibnamefont{Katsch}},
	\bibinfo{author}{\bibfnamefont{M.}~\bibnamefont{Selig}}, \bibnamefont{and}
	\bibinfo{author}{\bibfnamefont{A.}~\bibnamefont{Knorr}}, \bibinfo{journal}{2D
		Materials} \textbf{\bibinfo{volume}{7}}, \bibinfo{pages}{015021}
	(\bibinfo{year}{2020}).
	
	\bibitem[{\citenamefont{Kira and Koch}(2006)}]{kira2006many}
	\bibinfo{author}{\bibfnamefont{M.}~\bibnamefont{Kira}} \bibnamefont{and}
	\bibinfo{author}{\bibfnamefont{S.~W.} \bibnamefont{Koch}},
	\bibinfo{journal}{Progress in quantum electronics}
	\textbf{\bibinfo{volume}{30}}, \bibinfo{pages}{155} (\bibinfo{year}{2006}).
	
	\bibitem[{\citenamefont{Selig et~al.}(2016)\citenamefont{Selig, Bergh{\"a}user,
			Raja, Nagler, Sch{\"u}ller, Heinz, Korn, Chernikov, Mali\'{c}, and
			Knorr}}]{selig2016excitonic}
	\bibinfo{author}{\bibfnamefont{M.}~\bibnamefont{Selig}},
	\bibinfo{author}{\bibfnamefont{G.}~\bibnamefont{Bergh{\"a}user}},
	\bibinfo{author}{\bibfnamefont{A.}~\bibnamefont{Raja}},
	\bibinfo{author}{\bibfnamefont{P.}~\bibnamefont{Nagler}},
	\bibinfo{author}{\bibfnamefont{C.}~\bibnamefont{Sch{\"u}ller}},
	\bibinfo{author}{\bibfnamefont{T.~F.} \bibnamefont{Heinz}},
	\bibinfo{author}{\bibfnamefont{T.}~\bibnamefont{Korn}},
	\bibinfo{author}{\bibfnamefont{A.}~\bibnamefont{Chernikov}},
	\bibinfo{author}{\bibfnamefont{E.}~\bibnamefont{Mali\'{c}}},
	\bibnamefont{and} \bibinfo{author}{\bibfnamefont{A.}~\bibnamefont{Knorr}},
	\bibinfo{journal}{Nature Communications} \textbf{\bibinfo{volume}{7}},
	\bibinfo{pages}{13279} (\bibinfo{year}{2016}).
	
	\bibitem[{\citenamefont{Xiao et~al.}(2012)\citenamefont{Xiao, Liu, Feng, Xu,
			and Yao}}]{xiao2012coupled}
	\bibinfo{author}{\bibfnamefont{D.}~\bibnamefont{Xiao}},
	\bibinfo{author}{\bibfnamefont{G.-B.} \bibnamefont{Liu}},
	\bibinfo{author}{\bibfnamefont{W.}~\bibnamefont{Feng}},
	\bibinfo{author}{\bibfnamefont{X.}~\bibnamefont{Xu}}, \bibnamefont{and}
	\bibinfo{author}{\bibfnamefont{W.}~\bibnamefont{Yao}},
	\bibinfo{journal}{Physical Review Letters} \textbf{\bibinfo{volume}{108}},
	\bibinfo{pages}{196802} (\bibinfo{year}{2012}).
	
	\bibitem[{\citenamefont{Knorr et~al.}(1996)\citenamefont{Knorr, Hughes,
			Stroucken, and Koch}}]{knorr1996theory}
	\bibinfo{author}{\bibfnamefont{A.}~\bibnamefont{Knorr}},
	\bibinfo{author}{\bibfnamefont{S.}~\bibnamefont{Hughes}},
	\bibinfo{author}{\bibfnamefont{T.}~\bibnamefont{Stroucken}},
	\bibnamefont{and} \bibinfo{author}{\bibfnamefont{S.~W.} \bibnamefont{Koch}},
	\bibinfo{journal}{Chemical physics} \textbf{\bibinfo{volume}{210}},
	\bibinfo{pages}{27} (\bibinfo{year}{1996}).
	
	\bibitem[{\citenamefont{Axt and Stahl}(1994)}]{axt1994dynamics}
	\bibinfo{author}{\bibfnamefont{V.}~\bibnamefont{Axt}} \bibnamefont{and}
	\bibinfo{author}{\bibfnamefont{A.}~\bibnamefont{Stahl}},
	\bibinfo{journal}{Zeitschrift f{\"u}r Physik B Condensed Matter}
	\textbf{\bibinfo{volume}{93}}, \bibinfo{pages}{195} (\bibinfo{year}{1994}).
	
	\bibitem[{gam()}]{gamma}
	\bibinfo{note}{As biexcitons and continuous exciton-exciton scattering states
		are Coulomb generated by two exciton transitions, a radiative part (as
		contributing to the exciton linewidth) does not occur in $\gamma_\text{xx}$.
		linewidth, which in addition to a phonon-mediated part $\gamma_0$ also
		includes a radiative contribution.}
	
	\bibitem[{\citenamefont{Haase et~al.}(2000)\citenamefont{Haase, Neukirch,
			Meinertz, Gutowski, Axt, Bartels, Stahl, N{\"u}rnberger, and
			Faschinger}}]{haase2000intensity}
	\bibinfo{author}{\bibfnamefont{B.}~\bibnamefont{Haase}},
	\bibinfo{author}{\bibfnamefont{U.}~\bibnamefont{Neukirch}},
	\bibinfo{author}{\bibfnamefont{J.}~\bibnamefont{Meinertz}},
	\bibinfo{author}{\bibfnamefont{J.}~\bibnamefont{Gutowski}},
	\bibinfo{author}{\bibfnamefont{V.~M.} \bibnamefont{Axt}},
	\bibinfo{author}{\bibfnamefont{G.}~\bibnamefont{Bartels}},
	\bibinfo{author}{\bibfnamefont{A.}~\bibnamefont{Stahl}},
	\bibinfo{author}{\bibfnamefont{J.}~\bibnamefont{N{\"u}rnberger}},
	\bibnamefont{and}
	\bibinfo{author}{\bibfnamefont{W.}~\bibnamefont{Faschinger}},
	\bibinfo{journal}{Journal of Crystal Growth} \textbf{\bibinfo{volume}{214}},
	\bibinfo{pages}{852} (\bibinfo{year}{2000}).
	
	\bibitem[{\citenamefont{Emmanuele et~al.}(2019)\citenamefont{Emmanuele, Sich,
			Kyriienko, Shahnazaryan, Withers, Catanzaro, Walker, Benimetskiy, Skolnick,
			Tartakovskii et~al.}}]{emmanuele2019highly}
	\bibinfo{author}{\bibfnamefont{R.}~\bibnamefont{Emmanuele}},
	\bibinfo{author}{\bibfnamefont{M.}~\bibnamefont{Sich}},
	\bibinfo{author}{\bibfnamefont{O.}~\bibnamefont{Kyriienko}},
	\bibinfo{author}{\bibfnamefont{V.}~\bibnamefont{Shahnazaryan}},
	\bibinfo{author}{\bibfnamefont{F.}~\bibnamefont{Withers}},
	\bibinfo{author}{\bibfnamefont{A.}~\bibnamefont{Catanzaro}},
	\bibinfo{author}{\bibfnamefont{P.}~\bibnamefont{Walker}},
	\bibinfo{author}{\bibfnamefont{F.}~\bibnamefont{Benimetskiy}},
	\bibinfo{author}{\bibfnamefont{M.}~\bibnamefont{Skolnick}},
	\bibinfo{author}{\bibfnamefont{A.}~\bibnamefont{Tartakovskii}},
	\bibnamefont{et~al.}, \bibinfo{journal}{arXiv preprint arXiv:1910.14636}
	(\bibinfo{year}{2019}).
	
	\bibitem[{blu()}]{blueshift}
	\bibinfo{note}{Note that in the incoherent limit of a thermalized electron-hole
		plasma after well above band edge excitation, pronounced band gap shrinkage
		can lead to a red shift \cite{steinhoff2014influence,erben2018excitation}.}
	
	\bibitem[{\citenamefont{Mai et~al.}(2014)\citenamefont{Mai, Semenov, Barrette,
			Yu, Jin, Cao, Kim, and Gundogdu}}]{mai2014exciton}
	\bibinfo{author}{\bibfnamefont{C.}~\bibnamefont{Mai}},
	\bibinfo{author}{\bibfnamefont{Y.~G.} \bibnamefont{Semenov}},
	\bibinfo{author}{\bibfnamefont{A.}~\bibnamefont{Barrette}},
	\bibinfo{author}{\bibfnamefont{Y.}~\bibnamefont{Yu}},
	\bibinfo{author}{\bibfnamefont{Z.}~\bibnamefont{Jin}},
	\bibinfo{author}{\bibfnamefont{L.}~\bibnamefont{Cao}},
	\bibinfo{author}{\bibfnamefont{K.~W.} \bibnamefont{Kim}}, \bibnamefont{and}
	\bibinfo{author}{\bibfnamefont{K.}~\bibnamefont{Gundogdu}},
	\bibinfo{journal}{Physical Review B} \textbf{\bibinfo{volume}{90}},
	\bibinfo{pages}{041414(R)} (\bibinfo{year}{2014}).
	
	\bibitem[{\citenamefont{Sie et~al.}(2015{\natexlab{b}})\citenamefont{Sie,
			McIver, Lee, Fu, Kong, and Gedik}}]{sie2015valley}
	\bibinfo{author}{\bibfnamefont{E.~J.} \bibnamefont{Sie}},
	\bibinfo{author}{\bibfnamefont{J.~W.} \bibnamefont{McIver}},
	\bibinfo{author}{\bibfnamefont{Y.-H.} \bibnamefont{Lee}},
	\bibinfo{author}{\bibfnamefont{L.}~\bibnamefont{Fu}},
	\bibinfo{author}{\bibfnamefont{J.}~\bibnamefont{Kong}}, \bibnamefont{and}
	\bibinfo{author}{\bibfnamefont{N.}~\bibnamefont{Gedik}},
	\bibinfo{journal}{Nature Materials} \textbf{\bibinfo{volume}{14}},
	\bibinfo{pages}{290} (\bibinfo{year}{2015}{\natexlab{b}}).
	
	\bibitem[{\citenamefont{Wang et~al.}(2015)\citenamefont{Wang, Luo, Yabushita,
			Wu, Kobayashi, Chen, and Li}}]{wang2015ultrafast}
	\bibinfo{author}{\bibfnamefont{Y.-T.} \bibnamefont{Wang}},
	\bibinfo{author}{\bibfnamefont{C.-W.} \bibnamefont{Luo}},
	\bibinfo{author}{\bibfnamefont{A.}~\bibnamefont{Yabushita}},
	\bibinfo{author}{\bibfnamefont{K.-H.} \bibnamefont{Wu}},
	\bibinfo{author}{\bibfnamefont{T.}~\bibnamefont{Kobayashi}},
	\bibinfo{author}{\bibfnamefont{C.-H.} \bibnamefont{Chen}}, \bibnamefont{and}
	\bibinfo{author}{\bibfnamefont{L.-J.} \bibnamefont{Li}},
	\bibinfo{journal}{Scientific Reports} \textbf{\bibinfo{volume}{5}},
	\bibinfo{pages}{8289} (\bibinfo{year}{2015}).
	
	\bibitem[{\citenamefont{Sie et~al.}(2017)\citenamefont{Sie, Lui, Lee, Fu, Kong,
			and Gedik}}]{sie2017large}
	\bibinfo{author}{\bibfnamefont{E.~J.} \bibnamefont{Sie}},
	\bibinfo{author}{\bibfnamefont{C.~H.} \bibnamefont{Lui}},
	\bibinfo{author}{\bibfnamefont{Y.-H.} \bibnamefont{Lee}},
	\bibinfo{author}{\bibfnamefont{L.}~\bibnamefont{Fu}},
	\bibinfo{author}{\bibfnamefont{J.}~\bibnamefont{Kong}}, \bibnamefont{and}
	\bibinfo{author}{\bibfnamefont{N.}~\bibnamefont{Gedik}},
	\bibinfo{journal}{Science} \textbf{\bibinfo{volume}{355}},
	\bibinfo{pages}{1066} (\bibinfo{year}{2017}).
	
	\bibitem[{\citenamefont{Yong et~al.}(2018)\citenamefont{Yong, Horng, Shen, Cai,
			Wang, Yang, Lin, Zhao, Watanabe, Taniguchi et~al.}}]{yong2018biexcitonic}
	\bibinfo{author}{\bibfnamefont{C.-K.} \bibnamefont{Yong}},
	\bibinfo{author}{\bibfnamefont{J.}~\bibnamefont{Horng}},
	\bibinfo{author}{\bibfnamefont{Y.}~\bibnamefont{Shen}},
	\bibinfo{author}{\bibfnamefont{H.}~\bibnamefont{Cai}},
	\bibinfo{author}{\bibfnamefont{A.}~\bibnamefont{Wang}},
	\bibinfo{author}{\bibfnamefont{C.-S.} \bibnamefont{Yang}},
	\bibinfo{author}{\bibfnamefont{C.-K.} \bibnamefont{Lin}},
	\bibinfo{author}{\bibfnamefont{S.}~\bibnamefont{Zhao}},
	\bibinfo{author}{\bibfnamefont{K.}~\bibnamefont{Watanabe}},
	\bibinfo{author}{\bibfnamefont{T.}~\bibnamefont{Taniguchi}},
	\bibnamefont{et~al.}, \bibinfo{journal}{Nature Physics}
	\textbf{\bibinfo{volume}{14}}, \bibinfo{pages}{1092} (\bibinfo{year}{2018}).
	
	\bibitem[{\citenamefont{Christiansen et~al.}(2017)\citenamefont{Christiansen,
			Selig, Bergh\"auser, Schmidt, Niehues, Schneider, Arora, de~Vasconcellos,
			Bratschitsch, Mali\'{c} et~al.}}]{christiansen2017phonon}
	\bibinfo{author}{\bibfnamefont{D.}~\bibnamefont{Christiansen}},
	\bibinfo{author}{\bibfnamefont{M.}~\bibnamefont{Selig}},
	\bibinfo{author}{\bibfnamefont{G.}~\bibnamefont{Bergh\"auser}},
	\bibinfo{author}{\bibfnamefont{R.}~\bibnamefont{Schmidt}},
	\bibinfo{author}{\bibfnamefont{I.}~\bibnamefont{Niehues}},
	\bibinfo{author}{\bibfnamefont{R.}~\bibnamefont{Schneider}},
	\bibinfo{author}{\bibfnamefont{A.}~\bibnamefont{Arora}},
	\bibinfo{author}{\bibfnamefont{S.~M.} \bibnamefont{de~Vasconcellos}},
	\bibinfo{author}{\bibfnamefont{R.}~\bibnamefont{Bratschitsch}},
	\bibinfo{author}{\bibfnamefont{E.}~\bibnamefont{Mali\'{c}}},
	\bibnamefont{et~al.}, \bibinfo{journal}{Physical Review Letters}
	\textbf{\bibinfo{volume}{119}}, \bibinfo{pages}{187402}
	(\bibinfo{year}{2017}).
	
	\bibitem[{\citenamefont{Shree et~al.}(2018)\citenamefont{Shree, Semina, Robert,
			Han, Amand, Balocchi, Manca, Courtade, Marie, Taniguchi
			et~al.}}]{shree2018observation}
	\bibinfo{author}{\bibfnamefont{S.}~\bibnamefont{Shree}},
	\bibinfo{author}{\bibfnamefont{M.}~\bibnamefont{Semina}},
	\bibinfo{author}{\bibfnamefont{C.}~\bibnamefont{Robert}},
	\bibinfo{author}{\bibfnamefont{B.}~\bibnamefont{Han}},
	\bibinfo{author}{\bibfnamefont{T.}~\bibnamefont{Amand}},
	\bibinfo{author}{\bibfnamefont{A.}~\bibnamefont{Balocchi}},
	\bibinfo{author}{\bibfnamefont{M.}~\bibnamefont{Manca}},
	\bibinfo{author}{\bibfnamefont{E.}~\bibnamefont{Courtade}},
	\bibinfo{author}{\bibfnamefont{X.}~\bibnamefont{Marie}},
	\bibinfo{author}{\bibfnamefont{T.}~\bibnamefont{Taniguchi}},
	\bibnamefont{et~al.}, \bibinfo{journal}{Physical Review B}
	\textbf{\bibinfo{volume}{98}}, \bibinfo{pages}{035302}
	(\bibinfo{year}{2018}).
	
	\bibitem[{\citenamefont{Lengers et~al.}(2020)\citenamefont{Lengers, Kuhn, and
			Reiter}}]{lengers2020theory}
	\bibinfo{author}{\bibfnamefont{F.}~\bibnamefont{Lengers}},
	\bibinfo{author}{\bibfnamefont{T.}~\bibnamefont{Kuhn}}, \bibnamefont{and}
	\bibinfo{author}{\bibfnamefont{D.~E.} \bibnamefont{Reiter}},
	\bibinfo{journal}{Physical Review B} \textbf{\bibinfo{volume}{101}},
	\bibinfo{pages}{155304} (\bibinfo{year}{2020}).
	
	\bibitem[{\citenamefont{Schmidt et~al.}(2016)\citenamefont{Schmidt,
			Bergh{\"a}user, Schneider, Selig, Tonndorf, Mali\'{c}, Knorr, Michaelis~de
			Vasconcellos, and Bratschitsch}}]{schmidt2016ultrafast}
	\bibinfo{author}{\bibfnamefont{R.}~\bibnamefont{Schmidt}},
	\bibinfo{author}{\bibfnamefont{G.}~\bibnamefont{Bergh{\"a}user}},
	\bibinfo{author}{\bibfnamefont{R.}~\bibnamefont{Schneider}},
	\bibinfo{author}{\bibfnamefont{M.}~\bibnamefont{Selig}},
	\bibinfo{author}{\bibfnamefont{P.}~\bibnamefont{Tonndorf}},
	\bibinfo{author}{\bibfnamefont{E.}~\bibnamefont{Mali\'{c}}},
	\bibinfo{author}{\bibfnamefont{A.}~\bibnamefont{Knorr}},
	\bibinfo{author}{\bibfnamefont{S.}~\bibnamefont{Michaelis~de Vasconcellos}},
	\bibnamefont{and}
	\bibinfo{author}{\bibfnamefont{R.}~\bibnamefont{Bratschitsch}},
	\bibinfo{journal}{Nano Letters} \textbf{\bibinfo{volume}{16}},
	\bibinfo{pages}{2945} (\bibinfo{year}{2016}).
	
	\bibitem[{\citenamefont{Selig et~al.}(2018)\citenamefont{Selig, Bergh{\"a}user,
			Richter, Bratschitsch, Knorr, and Mali\'{c}}}]{selig2017dark}
	\bibinfo{author}{\bibfnamefont{M.}~\bibnamefont{Selig}},
	\bibinfo{author}{\bibfnamefont{G.}~\bibnamefont{Bergh{\"a}user}},
	\bibinfo{author}{\bibfnamefont{M.}~\bibnamefont{Richter}},
	\bibinfo{author}{\bibfnamefont{R.}~\bibnamefont{Bratschitsch}},
	\bibinfo{author}{\bibfnamefont{A.}~\bibnamefont{Knorr}}, \bibnamefont{and}
	\bibinfo{author}{\bibfnamefont{E.}~\bibnamefont{Mali\'{c}}},
	\bibinfo{journal}{2D Materials} \textbf{\bibinfo{volume}{5}},
	\bibinfo{pages}{035017} (\bibinfo{year}{2018}).
	
	\bibitem[{\citenamefont{Selig et~al.}(2019)\citenamefont{Selig, Katsch,
			Schmidt, de~Vasconcellos, Bratschitsch, Mali\'{c}, and
			Knorr}}]{selig2019ultrafast}
	\bibinfo{author}{\bibfnamefont{M.}~\bibnamefont{Selig}},
	\bibinfo{author}{\bibfnamefont{F.}~\bibnamefont{Katsch}},
	\bibinfo{author}{\bibfnamefont{R.}~\bibnamefont{Schmidt}},
	\bibinfo{author}{\bibfnamefont{S.~M.} \bibnamefont{de~Vasconcellos}},
	\bibinfo{author}{\bibfnamefont{R.}~\bibnamefont{Bratschitsch}},
	\bibinfo{author}{\bibfnamefont{E.}~\bibnamefont{Mali\'{c}}},
	\bibnamefont{and} \bibinfo{author}{\bibfnamefont{A.}~\bibnamefont{Knorr}},
	\bibinfo{journal}{Physical Review Research} \textbf{\bibinfo{volume}{1}},
	\bibinfo{pages}{022007(R)} (\bibinfo{year}{2019}).
	
	\bibitem[{\citenamefont{Christiansen et~al.}(2019)\citenamefont{Christiansen,
			Selig, Mali\'{c}, Ernstorfer, and Knorr}}]{christiansen2019theory}
	\bibinfo{author}{\bibfnamefont{D.}~\bibnamefont{Christiansen}},
	\bibinfo{author}{\bibfnamefont{M.}~\bibnamefont{Selig}},
	\bibinfo{author}{\bibfnamefont{E.}~\bibnamefont{Mali\'{c}}},
	\bibinfo{author}{\bibfnamefont{R.}~\bibnamefont{Ernstorfer}},
	\bibnamefont{and} \bibinfo{author}{\bibfnamefont{A.}~\bibnamefont{Knorr}},
	\bibinfo{journal}{Physical Review B} \textbf{\bibinfo{volume}{100}},
	\bibinfo{pages}{205401} (\bibinfo{year}{2019}).
	
	\bibitem[{\citenamefont{Selig et~al.}(2020)\citenamefont{Selig, Katsch, Brem,
			Mkrtchian, Malic, and Knorr}}]{selig2019quenching}
	\bibinfo{author}{\bibfnamefont{M.}~\bibnamefont{Selig}},
	\bibinfo{author}{\bibfnamefont{F.}~\bibnamefont{Katsch}},
	\bibinfo{author}{\bibfnamefont{S.}~\bibnamefont{Brem}},
	\bibinfo{author}{\bibfnamefont{G.~F.} \bibnamefont{Mkrtchian}},
	\bibinfo{author}{\bibfnamefont{E.}~\bibnamefont{Malic}}, \bibnamefont{and}
	\bibinfo{author}{\bibfnamefont{A.}~\bibnamefont{Knorr}},
	\bibinfo{journal}{Physical Review Research} \textbf{\bibinfo{volume}{2}},
	\bibinfo{pages}{023322} (\bibinfo{year}{2020}).
	
	\bibitem[{\citenamefont{Korm{\'a}nyos et~al.}(2015)\citenamefont{Korm{\'a}nyos,
			Burkard, Gmitra, Fabian, Z{\'o}lyomi, Drummond, and
			Fal’ko}}]{kormanyos2015k}
	\bibinfo{author}{\bibfnamefont{A.}~\bibnamefont{Korm{\'a}nyos}},
	\bibinfo{author}{\bibfnamefont{G.}~\bibnamefont{Burkard}},
	\bibinfo{author}{\bibfnamefont{M.}~\bibnamefont{Gmitra}},
	\bibinfo{author}{\bibfnamefont{J.}~\bibnamefont{Fabian}},
	\bibinfo{author}{\bibfnamefont{V.}~\bibnamefont{Z{\'o}lyomi}},
	\bibinfo{author}{\bibfnamefont{N.~D.} \bibnamefont{Drummond}},
	\bibnamefont{and} \bibinfo{author}{\bibfnamefont{V.}~\bibnamefont{Fal’ko}},
	\bibinfo{journal}{2D Materials} \textbf{\bibinfo{volume}{2}},
	\bibinfo{pages}{022001} (\bibinfo{year}{2015}).
	
	\bibitem[{\citenamefont{Taniguchi and Watanabe}(2007)}]{taniguchi2007synthesis}
	\bibinfo{author}{\bibfnamefont{T.}~\bibnamefont{Taniguchi}} \bibnamefont{and}
	\bibinfo{author}{\bibfnamefont{K.}~\bibnamefont{Watanabe}},
	\bibinfo{journal}{Journal of Crystal Growth} \textbf{\bibinfo{volume}{303}},
	\bibinfo{pages}{525} (\bibinfo{year}{2007}).
	
	\bibitem[{\citenamefont{Afrousheh et~al.}(2004)\citenamefont{Afrousheh,
			Bohlouli-Zanjani, Vagale, Mugford, Fedorov, and
			Martin}}]{afrousheh2004spectroscopic}
	\bibinfo{author}{\bibfnamefont{K.}~\bibnamefont{Afrousheh}},
	\bibinfo{author}{\bibfnamefont{P.}~\bibnamefont{Bohlouli-Zanjani}},
	\bibinfo{author}{\bibfnamefont{D.}~\bibnamefont{Vagale}},
	\bibinfo{author}{\bibfnamefont{A.}~\bibnamefont{Mugford}},
	\bibinfo{author}{\bibfnamefont{M.}~\bibnamefont{Fedorov}}, \bibnamefont{and}
	\bibinfo{author}{\bibfnamefont{J.~D.~D.} \bibnamefont{Martin}},
	\bibinfo{journal}{Physical Review Letters} \textbf{\bibinfo{volume}{93}},
	\bibinfo{pages}{233001} (\bibinfo{year}{2004}).
	
	\bibitem[{\citenamefont{Liu et~al.}(2014)\citenamefont{Liu, Neal, Zhu, Luo, Xu,
			Tom{\'a}nek, and Ye}}]{liu2014phosphorene}
	\bibinfo{author}{\bibfnamefont{H.}~\bibnamefont{Liu}},
	\bibinfo{author}{\bibfnamefont{A.~T.} \bibnamefont{Neal}},
	\bibinfo{author}{\bibfnamefont{Z.}~\bibnamefont{Zhu}},
	\bibinfo{author}{\bibfnamefont{Z.}~\bibnamefont{Luo}},
	\bibinfo{author}{\bibfnamefont{X.}~\bibnamefont{Xu}},
	\bibinfo{author}{\bibfnamefont{D.}~\bibnamefont{Tom{\'a}nek}},
	\bibnamefont{and} \bibinfo{author}{\bibfnamefont{P.~D.} \bibnamefont{Ye}},
	\bibinfo{journal}{ACS Nano} \textbf{\bibinfo{volume}{8}},
	\bibinfo{pages}{4033} (\bibinfo{year}{2014}).
	
	\bibitem[{\citenamefont{Tran et~al.}(2014)\citenamefont{Tran, Soklaski, Liang,
			and Yang}}]{tran2014layer}
	\bibinfo{author}{\bibfnamefont{V.}~\bibnamefont{Tran}},
	\bibinfo{author}{\bibfnamefont{R.}~\bibnamefont{Soklaski}},
	\bibinfo{author}{\bibfnamefont{Y.}~\bibnamefont{Liang}}, \bibnamefont{and}
	\bibinfo{author}{\bibfnamefont{L.}~\bibnamefont{Yang}},
	\bibinfo{journal}{Physical Review B} \textbf{\bibinfo{volume}{89}},
	\bibinfo{pages}{235319} (\bibinfo{year}{2014}).
	
	\bibitem[{\citenamefont{Li et~al.}(2014)\citenamefont{Li, Yu, Ye, Ge, Ou, Wu,
			Feng, Chen, and Zhang}}]{li2014black}
	\bibinfo{author}{\bibfnamefont{L.}~\bibnamefont{Li}},
	\bibinfo{author}{\bibfnamefont{Y.}~\bibnamefont{Yu}},
	\bibinfo{author}{\bibfnamefont{G.~J.} \bibnamefont{Ye}},
	\bibinfo{author}{\bibfnamefont{Q.}~\bibnamefont{Ge}},
	\bibinfo{author}{\bibfnamefont{X.}~\bibnamefont{Ou}},
	\bibinfo{author}{\bibfnamefont{H.}~\bibnamefont{Wu}},
	\bibinfo{author}{\bibfnamefont{D.}~\bibnamefont{Feng}},
	\bibinfo{author}{\bibfnamefont{X.~H.} \bibnamefont{Chen}}, \bibnamefont{and}
	\bibinfo{author}{\bibfnamefont{Y.}~\bibnamefont{Zhang}},
	\bibinfo{journal}{Nature Nanotechnology} \textbf{\bibinfo{volume}{9}},
	\bibinfo{pages}{372} (\bibinfo{year}{2014}).
	
	\bibitem[{\citenamefont{Rudenko and
			Katsnelson}(2014)}]{rudenko2014quasiparticle}
	\bibinfo{author}{\bibfnamefont{A.~N.} \bibnamefont{Rudenko}} \bibnamefont{and}
	\bibinfo{author}{\bibfnamefont{M.~I.} \bibnamefont{Katsnelson}},
	\bibinfo{journal}{Physical Review B} \textbf{\bibinfo{volume}{89}},
	\bibinfo{pages}{201408(R)} (\bibinfo{year}{2014}).
	
	\bibitem[{\citenamefont{Deilmann and Thygesen}(2018)}]{deilmann2018unraveling}
	\bibinfo{author}{\bibfnamefont{T.}~\bibnamefont{Deilmann}} \bibnamefont{and}
	\bibinfo{author}{\bibfnamefont{K.~S.} \bibnamefont{Thygesen}},
	\bibinfo{journal}{2D Materials} \textbf{\bibinfo{volume}{5}},
	\bibinfo{pages}{041007} (\bibinfo{year}{2018}).
	
	\bibitem[{\citenamefont{Dai and Zeng}(2015)}]{dai2015titanium}
	\bibinfo{author}{\bibfnamefont{J.}~\bibnamefont{Dai}} \bibnamefont{and}
	\bibinfo{author}{\bibfnamefont{X.~C.} \bibnamefont{Zeng}},
	\bibinfo{journal}{Angewandte Chemie International Edition}
	\textbf{\bibinfo{volume}{54}}, \bibinfo{pages}{7572} (\bibinfo{year}{2015}).
	
	\bibitem[{\citenamefont{Van~der Donck and Peeters}(2018)}]{van2018excitonic}
	\bibinfo{author}{\bibfnamefont{M.}~\bibnamefont{Van~der Donck}}
	\bibnamefont{and} \bibinfo{author}{\bibfnamefont{F.~M.}
		\bibnamefont{Peeters}}, \bibinfo{journal}{Physical Review B}
	\textbf{\bibinfo{volume}{98}}, \bibinfo{pages}{235401}
	(\bibinfo{year}{2018}).
	
	\bibitem[{\citenamefont{Cassette et~al.}(2015)\citenamefont{Cassette, Pensack,
			Mahler, and Scholes}}]{cassette2015room}
	\bibinfo{author}{\bibfnamefont{E.}~\bibnamefont{Cassette}},
	\bibinfo{author}{\bibfnamefont{R.~D.} \bibnamefont{Pensack}},
	\bibinfo{author}{\bibfnamefont{B.}~\bibnamefont{Mahler}}, \bibnamefont{and}
	\bibinfo{author}{\bibfnamefont{G.~D.} \bibnamefont{Scholes}},
	\bibinfo{journal}{Nature Communications} \textbf{\bibinfo{volume}{6}},
	\bibinfo{pages}{6086} (\bibinfo{year}{2015}).
	
	\bibitem[{\citenamefont{Wu et~al.}(2015)\citenamefont{Wu, Trinh, and
			Zhu}}]{wu2015excitonic}
	\bibinfo{author}{\bibfnamefont{X.}~\bibnamefont{Wu}},
	\bibinfo{author}{\bibfnamefont{M.~T.} \bibnamefont{Trinh}}, \bibnamefont{and}
	\bibinfo{author}{\bibfnamefont{X.-Y.} \bibnamefont{Zhu}},
	\bibinfo{journal}{The Journal of Physical Chemistry C}
	\textbf{\bibinfo{volume}{119}}, \bibinfo{pages}{14714}
	(\bibinfo{year}{2015}).
	
	\bibitem[{\citenamefont{Thouin et~al.}(2019)\citenamefont{Thouin, Cortecchia,
			Petrozza, Srimath~Kandada, and Silva}}]{thouin2019enhanced}
	\bibinfo{author}{\bibfnamefont{F.}~\bibnamefont{Thouin}},
	\bibinfo{author}{\bibfnamefont{D.}~\bibnamefont{Cortecchia}},
	\bibinfo{author}{\bibfnamefont{A.}~\bibnamefont{Petrozza}},
	\bibinfo{author}{\bibfnamefont{A.~R.} \bibnamefont{Srimath~Kandada}},
	\bibnamefont{and} \bibinfo{author}{\bibfnamefont{C.}~\bibnamefont{Silva}},
	\bibinfo{journal}{Physical Review Research} \textbf{\bibinfo{volume}{1}},
	\bibinfo{pages}{032032} (\bibinfo{year}{2019}).
	
	\bibitem[{\citenamefont{Axt and Mukamel}(1998)}]{axt1998nonlinear}
	\bibinfo{author}{\bibfnamefont{V.~M.} \bibnamefont{Axt}} \bibnamefont{and}
	\bibinfo{author}{\bibfnamefont{S.}~\bibnamefont{Mukamel}},
	\bibinfo{journal}{Reviews of Modern Physics} \textbf{\bibinfo{volume}{70}},
	\bibinfo{pages}{145} (\bibinfo{year}{1998}).
	
	\bibitem[{\citenamefont{Agranovich et~al.}(1998)\citenamefont{Agranovich,
			Basko, La~Rocca, and Bassani}}]{agranovich1998excitons}
	\bibinfo{author}{\bibfnamefont{V.~M.} \bibnamefont{Agranovich}},
	\bibinfo{author}{\bibfnamefont{D.~M.} \bibnamefont{Basko}},
	\bibinfo{author}{\bibfnamefont{G.~C.} \bibnamefont{La~Rocca}},
	\bibnamefont{and} \bibinfo{author}{\bibfnamefont{F.}~\bibnamefont{Bassani}},
	\bibinfo{journal}{Journal of Physics: Condensed Matter}
	\textbf{\bibinfo{volume}{10}}, \bibinfo{pages}{9369} (\bibinfo{year}{1998}).
	
	\bibitem[{\citenamefont{Abdel-Baki et~al.}(2016)\citenamefont{Abdel-Baki,
			Boitier, Diab, Lanty, Jemli, L{\'e}d{\'e}e, Garrot, Deleporte, and
			Lauret}}]{abdel2016exciton}
	\bibinfo{author}{\bibfnamefont{K.}~\bibnamefont{Abdel-Baki}},
	\bibinfo{author}{\bibfnamefont{F.}~\bibnamefont{Boitier}},
	\bibinfo{author}{\bibfnamefont{H.}~\bibnamefont{Diab}},
	\bibinfo{author}{\bibfnamefont{G.}~\bibnamefont{Lanty}},
	\bibinfo{author}{\bibfnamefont{K.}~\bibnamefont{Jemli}},
	\bibinfo{author}{\bibfnamefont{F.}~\bibnamefont{L{\'e}d{\'e}e}},
	\bibinfo{author}{\bibfnamefont{D.}~\bibnamefont{Garrot}},
	\bibinfo{author}{\bibfnamefont{E.}~\bibnamefont{Deleporte}},
	\bibnamefont{and} \bibinfo{author}{\bibfnamefont{J.}~\bibnamefont{Lauret}},
	\bibinfo{journal}{Journal of Applied Physics} \textbf{\bibinfo{volume}{119}},
	\bibinfo{pages}{064301} (\bibinfo{year}{2016}).
	
	\bibitem[{\citenamefont{Steinhoff et~al.}(2014)\citenamefont{Steinhoff, Rosner,
			Jahnke, Wehling, and Gies}}]{steinhoff2014influence}
	\bibinfo{author}{\bibfnamefont{A.}~\bibnamefont{Steinhoff}},
	\bibinfo{author}{\bibfnamefont{M.}~\bibnamefont{Rosner}},
	\bibinfo{author}{\bibfnamefont{F.}~\bibnamefont{Jahnke}},
	\bibinfo{author}{\bibfnamefont{T.~O.} \bibnamefont{Wehling}},
	\bibnamefont{and} \bibinfo{author}{\bibfnamefont{C.}~\bibnamefont{Gies}},
	\bibinfo{journal}{Nano Letters} \textbf{\bibinfo{volume}{14}},
	\bibinfo{pages}{3743} (\bibinfo{year}{2014}).
	
	\bibitem[{\citenamefont{Erben et~al.}(2018)\citenamefont{Erben, Steinhoff,
			Gies, Sch{\"o}nhoff, Wehling, and Jahnke}}]{erben2018excitation}
	\bibinfo{author}{\bibfnamefont{D.}~\bibnamefont{Erben}},
	\bibinfo{author}{\bibfnamefont{A.}~\bibnamefont{Steinhoff}},
	\bibinfo{author}{\bibfnamefont{C.}~\bibnamefont{Gies}},
	\bibinfo{author}{\bibfnamefont{G.}~\bibnamefont{Sch{\"o}nhoff}},
	\bibinfo{author}{\bibfnamefont{T.~O.} \bibnamefont{Wehling}},
	\bibnamefont{and} \bibinfo{author}{\bibfnamefont{F.}~\bibnamefont{Jahnke}},
	\bibinfo{journal}{Physical Review B} \textbf{\bibinfo{volume}{98}},
	\bibinfo{pages}{035434} (\bibinfo{year}{2018}).
	
\end{thebibliography}
\end{document}